\documentclass[3p]{elsarticle}

\usepackage{graphicx}
\graphicspath{{figures/}}
\usepackage{caption}
\usepackage{subcaption}
\usepackage{pdfpages}
\usepackage{amsmath}  
\usepackage{nicefrac}
\usepackage{amssymb}          
\usepackage{booktabs}
\usepackage{multirow}
\usepackage{color, colortbl}
\usepackage{lineno,hyperref}
\modulolinenumbers[5]
\hypersetup{pdfauthor=author}

\makeatletter
\newcommand{\setword}[2]{%
  \phantomsection
  #1\def\@currentlabel{\unexpanded{#1}}\label{#2}%
}
\makeatother

\definecolor{DynCyan}{rgb}{0.1058, 0.4549, 0.5686} % 27, 116, 145 #1b7491
\definecolor{DynRed}{rgb}{0.5529, 0.1215, 0.1333} % 141, 31, 34 #8d1f22
\definecolor{DynGrey}{rgb}{0.3568, 0.4509, 0.5098} % 91, 115, 130 #5b7382
\definecolor{DynDark}{rgb}{0.0470, 0.2000, 0.2470} % 12, 51, 63 #0c333f

\journal{Ocean Engineering}

\biboptions{authoryear}

\begin{document}

\begin{frontmatter}

\title{Machine learning for phase-resolved reconstruction of nonlinear ocean wave surface elevations from sparse remote sensing data}

\author[DYN]{Svenja Ehlers\corref{mycorrespondingauthor}}
\ead{svenja.ehlers@tuhh.de} % \ead[url]{www.tuhh.de/dyn}

\author[DLR,DYN]{Marco Klein}
\author[TUD]{Alexander Heinlein}
\author[DYN]{Mathies Wedler}
\author[DLR,DYN]{Nicolas Desmars}
\author[DYN,ICL]{Norbert Hoffmann}
\author[TUB]{Merten Stender}

\cortext[mycorrespondingauthor]{Corresponding author}

% \affiliation[DYN]{
%     organization={Hamburg University of Technology, Dynamics Group},
%     addressline={Schloßmühlendamm 30},
%     postcode={21073},
%     city={Hamburg},
%     country={Germany}
% }
% \affiliation[DLR]{
%     organization={German Aerospace Center (DLR), Institute of Maritime Energy Systems},
%     postcode={21502},
%     city={Geesthacht},
%     country={Germany}
% }
% \affiliation[TUD]{
%     organization={Delft University of Technology, Delft Institute of Applied Mathematics },
%     postcode={2628 CD},
%     city={Delft},
%     country={Netherlands}
% }
% \affiliation[ICL]{
%     organization={Imperial College London, Department of Mechanical Engineering},
%     city={London SW7 2AZ},
%     country={United Kingdom}
% }
% \affiliation[TUB]{
%     organization={Technische Universität Berlin, Cyber-Physical Systems in Mechanical Engineering},
%     postcode={10623},
%     city={Berlin},
%     country={Germany}
% }

\address[DYN]{Hamburg University of Technology, Dynamics Group, Schloßmühlendamm 30, 21073 Hamburg, Germany}
\address[DLR]{German Aerospace Center, Institute of Maritime Energy Systems, Ship Performance Dep., 21502 Geesthacht, Germany}
\address[TUD]{Delft University of Technology, Delft Institute of Applied Mathematics, 2628 CD Delft, Netherlands}
\address[ICL]{Imperial College London, Department of Mechanical Engineering, London SW7 2AZ, United Kingdom}
\address[TUB]{Technische Universität Berlin, Cyber-Physical Systems in Mechanical Engineering, 10623 Berlin, Germany}

\begin{abstract}
Accurate short-term predictions of phase-resolved water wave conditions are crucial for decision-making in ocean engineering. However, the initialization of remote-sensing-based wave prediction models first requires a reconstruction of wave surfaces from sparse measurements like radar. Existing reconstruction methods either rely on computationally intensive optimization procedures or simplistic modelling assumptions that compromise the real-time capability or accuracy of the subsequent prediction process. We therefore address these issues by proposing a novel approach for phase-resolved wave surface reconstruction using neural networks based on the U-Net and Fourier neural operator (FNO) architectures. Our approach utilizes synthetic yet highly realistic training data on uniform one-dimensional grids, that is generated by the high-order spectral method for wave simulation and a geometric radar modelling approach. The investigation reveals that both models deliver accurate wave reconstruction results and show good generalization for different sea states when trained with spatio-temporal radar data containing multiple historic radar snapshots in each input. Notably, the FNO demonstrates superior performance in handling the data structure imposed by wave physics due to its global approach to learn the mapping between input and output in Fourier space.

\end{abstract}

% \begin{graphicalabstract}
% \includegraphics{figures/graphical_abstract.pdf}
% \end{graphicalabstract}

% \begin{highlights}
% \item fast phase-resolved inversion of ocean wave surfaces from radar measurements 
% \item application of machine learning via Fourier neural operator and U-Net based models
% \item Fourier neural operator found to be particularly suitable for wave data structure
% \item spatio-temporal radar data advantageous for high-accuracy wave reconstructions
% \end{highlights}


\end{frontmatter}

%%% Introduction %%%

\section{Introduction}

\begin{figure}[ht]
\centering
\begin{footnotesize}
\includegraphics{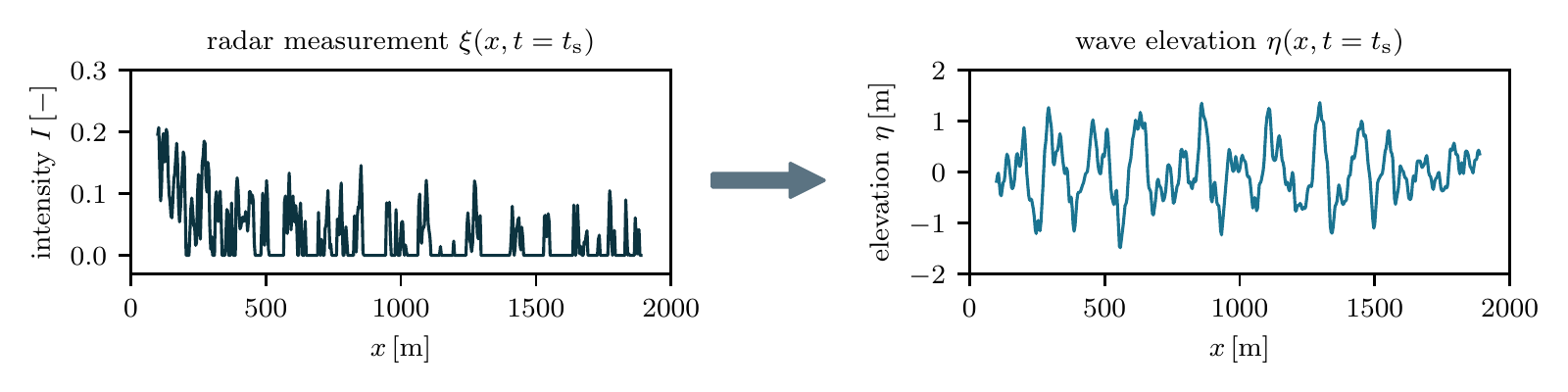}
\end{footnotesize}
\vspace{-0.4cm}
\caption{Graphical illustration of the phase-resolved reconstruction task of ocean wave surfaces $\eta$ from sparse radar intensity surfaces $\xi$ for the case of waves travelling in one spatial dimension. The radar measurement (left panel) is a snapshot acquired at time instant $t_\mathrm{s}$ and is considered as \textit{sparse} due to reoccurring areas with zero intensity caused by the geometrical shadowing modulation. This radar snapshot is used for reconstructing the wave surface elevation at the same time instant $t_\mathrm{s}$ (right panel).  }
\label{fig:graphical abstract}
\vspace{-0.4cm}
\end{figure}

Offshore installations and vessels are strongly impacted by the dynamics of the surrounding ocean waves. Thus, accurate predictions of future wave conditions are desirable for enhancing their safe and efficient operation. For this purpose, several numerical methods have been developed, involving two fundamental steps: the assimilation and reconstruction of initial wave conditions from wave measurement data, followed by the prediction of the future wave evolution.  While one line of research focuses on predicting simplified phase-averaged wave quantities based on statistical parameters, marine applications such as wind turbine installations, helicopter landings, or control of wave energy converters require phase-resolved spatio-temporal wave information $\eta(x,t)$ to identify periods of low wave conditions or enable extreme event warnings. The X-band radar is a remote sensing device that can obtain such phase-resolved wave information. However, the radar backscatter is affected by the geometrical mechanism of tilt and shadowing modulation, creating a nonlinear and sparse relationship between radar measurement intensities $\xi(x,t)$ and the actual ocean wave surface elevation $\eta(x,t)$. This makes a reconstruction of wave information from radar information necessary in the assimilation step, which is also referred to as \textit{radar inversion} and is graphically exemplified in Figure~\ref{fig:graphical abstract}.

Contemporary phase-resolved wave reconstruction and prediction methods face a trade-off between accuracy and real-time capability. To achieve computationally efficient methods, linear wave theory (LWT) is commonly employed during the prediction step \citep[cf.][]{Morris1998, Naaijen2014, Hilmer2015}, along with prior spectral- or texture-analysis-based reconstruction of initial wave conditions from radar data \citep{NietoBorge2004, Dankert2004}. However, these reconstruction methods necessitate additional calibration by wave buoys or rely on simplified assumptions concerning the radar backscatter. Furthermore, the accuracy of the linear approach decreases remarkably for larger temporal horizons of prediction and increasing wave steepness \citep{Luenser2022}, necessitating a wave prediction using nonlinear wave models, especially for capturing safety-critical events such as rogue waves \citep{Ducrozet2007, Kharif2009}. Comparative studies on phase-resolved nonlinear ocean wave prediction have demonstrated that the high-order spectral (HOS) method, introduced by \cite{West1987} and \cite{Dommermuth1987}, provides the best prediction accuracy over a wide spatio-temporal domain as well as characteristic wave steepness \citep{Klein2020, Wu2004, Luenser2022, Blondel2009}. While the HOS prediction step itself is also numerically efficient, the reconstruction step currently represents the weakest part in the entire process \citep{Koellisch2018}: the inversion of initial conditions relies on an optimization procedure of the wave model parameters for the subsequent prediction \citep{Wu2004, Blondel2009}, which decreases the possible horizon of prediction and hinders the real-time capability so far \citep{Desmars2020}. Even though the alternative for the HOS inversion proposed by \cite{Koellisch2018} is able to improve the real-time capability, this method instead assumes an unrealistic radar snapshots data rate $\Delta t_\mathrm{r}$, making it not suitable for real-world applications \citep{Desmars2020}.

The aforementioned shortcomings of conventional ocean wave reconstruction and prediction methods have motivated the exploration of alternatives based on machine learning (ML) techniques. For instance, ML methods are able to predict simple phase-averaged wave quantities such as significant wave height $H_\mathrm{s}$, peak period $T_\mathrm{p}$ or mean wave direction \citep[cf.][]{Deo2001, Asma2012, James2018, Wu2020, Yevnin2022}. Recent advancements have also allowed for the more complex task of predicting the spatio-temporal evolution of phase-resolved wave fields, achieved by training multilayer perceptrons (MLPs) \citep{Desouky2019, Law2020, Duan2020, Zhang2022}, recurrent neural networks (RNNs) \citep{Kagemoto2020, Mohaghegh2021, Liu2022}, or convolutional neural networks (CNNs) \citep{Klein2022, Wedler2023} on synthetic or experimental one-dimensional elevation data.
However, these studies presuppose that either temporal sequences of wave elevations can be solely measured at a single point in space by buoys $\eta(x=x_\mathrm{p}, t)$  or snapshots of initial wave conditions are available throughout the entire space domain $\eta(x, t=t_\mathrm{s})$.  In practice, neither of these assumptions is feasible due to the lack of directional wave information of single-point measurements and the fact that the acquisition of spatial snapshots using remote sensing systems such as radars leads to sparse and unscaled observations $\xi(x, t=t_\mathrm{s})$, requiring a reconstruction of wave surface elevations first.

Consequently, it would be advantageous to employ ML methods also for the phase-resolved reconstruction of wave elevations $\eta(x,t)$ from X-band radar data $\xi(x,t)$. However, as far as the authors are aware, this topic has not yet been addressed. Prior studies have solely focused on reconstructing phase-averaged statistical parameters of the prevailing sea state from radar data. For instance, \cite{Vicen-Bueno2012} and \cite{Salcedo-Sanz2015} improved the estimation of $H_\mathrm{s}$ by extracting scalar features from sequences of radar images $\xi(x,t)$ in a preprocessing step, which in turn were employed to train MLPs and support vector regression models. In contrast, \cite{yang2021} extracted features from each of the consecutive radar images itself for improved $H_s$ estimation at the current time instant. While these methods rely on handcrafted features acquired during a preprocessing step, end-to-end approaches that automatically extract important features from their input have also been proposed. For instance, \cite{Duan2020b} and \cite{Chen2022} used CNN-based methods to estimate $H_\mathrm{s}$ and $T_\mathrm{p}$ from radar images.\\

Although there seems to be no relevant research on ML-based reconstruction of phase-resolved wave surfaces from sparse X-Band radar data, we hypothesize that ML offers a valuable alternative for the radar inversion task (\setword{Hypothesis~1}{hyp:1}). This hypothesis is derived from the observation that the reconstruction of zero-valued areas in the radar input, exemplified in Figure~\ref{fig:graphical abstract}, shares similarities with typical inverse problems encountered in imaging \citep{Bertero2022, Ongie2020} such as inpainting and restoration, where ML-methods have demonstrated successful applications \citep{Pathak2016, Zhang2017}. Two neural network architectures, with network components involving either a local or global approach of data processing, are investigated in detail for their performance in our task. Specifically, we will adapt the U-Net proposed by \cite{Ronneberger2015}, a fully convolutional neural network that employs a mapping approach in Euclidean space, and the Fourier neural operator (FNO) proposed by \cite{Li2020}, which is designed to learn a more global mapping in Fourier space. Despite the success of CNN-based approaches in imaging problems, we hypothesize that FNO models may be better suited for handling the complex and dynamic nature of ocean waves (\setword{Hypothesis~2}{hyp:2}), since we can assume that the wave features are already explicitly encoded in the network structure, as it learns data patterns in Fourier space. In contrast, the U-Net needs to learn these wave features by aggregating information from multiple layers. Lastly, we expect that incorporating historical context via spatio-temporal radar data will enhance the reconstruction quality of both ML architectures (\setword{Hypothesis~3}{hyp:3}), which we infer from classical radar inversion methods that also rely on temporal sequences of multiple radar snapshots \citep[cf.][]{Dankert2004, NietoBorge2004}.

In general, the fast inference capabilities of trained ML models, make them ideal for maintaining the real-time capability of the entire process composed of wave reconstruction and prediction (\setword{Criterion 1}{cri:1}) due to the rapid surface reconstruction without particular data preprocessing. Besides the real-time capability, ensuring high reconstruction accuracy is crucial to prevent initial reconstruction errors that will accumulate and deteriorate the subsequent wave prediction. Hence, we strive for an empirical reference value for the surface similarity parameter (SSP) error metric \citep{Perlin2014} of less than $\mathrm{SSP} \le 0.10$ between ground truth and reconstructed wave surfaces (\setword{Criterion 2}{cri:2}), which is a commonly used error threshold in ocean wave research \citep{Klein2020, Luenser2022}. In addition, the proposed ML methods must be capable of handling real-world measurement conditions of radar snapshots taken at intervals of $\Delta t_\mathrm{r} = [1, \,2] \, \mathrm{s}$ (\setword{Criterion 3}{cri:3}), a common X-band radar revolution period \citep{Neill2017}.\\

To summarize, the objective of this work is to develop an ML-based approach for phase-resolved radar inversion. This involves training ML models to learn mapping functions $\mathcal{M}$ that are able to reconstruct spatial wave elevation snapshots $\eta(x, t=t_\mathrm{s})$ from one or $n_\mathrm{s}$ consecutive historical radar snapshots $\xi(x, t_j)$, where $t_j = \{ t_\mathrm{s}-j \Delta t_\mathrm{r} \}_{j=0, \dots, n_\mathrm{s}-1} $. As obtaining ground truth wave surface elevation data for large spatial domains in real ocean conditions is almost impractical, we first generate synthetic yet highly realistic one-dimensional spatio-temporal wave surfaces $\eta(x,t)$ using the HOS method for different sea states in Section~\ref{sec:data generation}. The corresponding X-band radar surfaces $\xi(x,t)$ are generated using a geometric approach and incorporate tilt- and shadowing modulations. In Section~\ref{sec:ML Methodolgy}, two neural network architectures are introduced, a U-Net-based and FNO-based network, which are investigated for their suitability for radar inversion. In Section~\ref{sec:results}, we discuss the computational results. In particular, we first compare the wave reconstruction performance of the U-Net-based and the FNO-based models, each trained using either $n_\mathrm{s}=1$ radar snapshot in each input or spatio-temporal input data, meaning that multiple consecutive radar snapshots $n_\mathrm{s}$ are provided. Afterwards, the observations are generalized for the entire data set and discussed. Finally, in Section~\ref{sec:conclusion}, we draw conclusions based on these results and suggest future research directions.

%%% Methods %%%
\section{Data generation and preparation}
\label{sec:data generation}

This section briefly introduces the generation of long-crested nonlinear synthetic wave data $\eta(x,t)$ using the HOS method, followed by the generation of synthetic radar data $\xi(x,t)$ that accounts for the tilt- and shadowing modulation mechanisms. The final step involves extracting a number of $N$ input-output $(\mathbf{x}_i$,$\mathbf{y}_i),i=1, \dots, N$  data samples from the synthetic radar and wave data, which  we employ to train the supervised ML models. This first study on ML-based phase-resolved wave reconstruction focuses on the scenario of one-dimensional wave and radar data, driven by the advantages of easier data generation, simplified implementation, and faster neural network training with fewer computational resources. 

\subsection{Nonlinear synthetic wave data}

To generate synthetic one-dimensional wave data, the water-wave problem can be expressed by potential flow theory. Assuming a Newtonian fluid that is incompressible, inviscid, and irrotational, the underlying wave model is described by a velocity potential $\Phi(x, z, t)$ satisfying the \textit{Laplace equation}
\begin{equation}
    \nabla^2 \Phi = \frac{\partial^2\Phi}{\partial x^2} + \frac{\partial^2\Phi}{\partial z^2}=0 \label{eq:Laplace}
\end{equation}
within the fluid domain, where $z=0 \, \mathrm{m}$ is the mean free surface with $z$ pointing in upward direction. The domain is bounded by the \textit{kinematic} and \textit{dynamic boundary conditions} at the free surface $\eta(x,t) $ and the \textit{bottom boundary condition} at the seabed at depth $d$
\begin{align}
    \eta_t +  \eta_x \Phi_x - \Phi_z =0     \hspace{1cm}  &   \text{on } z=\eta(x, t) \nonumber \\ 
    \Phi_t +g \eta   + \frac{1}{2} \left(\Phi_{xx}^2 +\Phi_{zz}^2  \right)=0 \hspace{1cm} &   \text{on } z=\eta(x, t)\\
    \Phi_z =0 \hspace{1cm} &   \text{on } z=-d.  \nonumber
\end{align}
Solving this system of equations is challenging due to the nonlinear terms in the boundary conditions, which must be satisfied additionally at the unknown free surface $\eta(x,t)$.  Even though linear wave theory \citep{Airy1845} provides adequate approximations for certain engineering applications, capturing realistic ocean wave effects requires modelling the nonlinear behaviour of surface gravity waves. Thus, we employ the HOS method, as formulated by \cite{West1987}, which transforms the boundary conditions to the free surface and expresses them as a perturbation series of nonlinear order $M$ around $z=0$. In practice, an order of $M \le 4$ is sufficient for capturing the nonlinear wave effects of interest \citep{Desmars2020, Luenser2022}. The HOS simulation is linearly initialized by spatial wave surface elevation snapshots $\eta(x, t_\mathrm{s}=0)$ sampled from the JONSWAP spectrum for finite water depth \citep{Hasselmann1973,  Bouws1985}. The corresponding initial potential is linearly approximated. Subsequently, the initial elevation and potential are propagated nonlinearly in time with the chosen HOS order $M$.
The referred JONSWAP spectrum attains its maximum at a peak frequency $\omega_\mathrm{p}$, whereas the peak enhancement factor $\gamma$ determines the energy distribution around $\omega_\mathrm{p}$. The wave frequencies $ \omega$ are linked to the wavenumbers $k$ by the linear dispersion relation $\omega = \sqrt{g k \cdot \tanh{(k d)}}$. The relations $\omega = \nicefrac{2 \pi}{T}$ and $k=\nicefrac{2\pi}{L}$ allow for substituting the peak frequency with a peak period $T_\mathrm{p}$, peak wavelength $L_\mathrm{p}$, or peak wavenumber $k_\mathrm{p}$. Moreover, a dimensionless wave steepness parameter $\epsilon = k_\mathrm{p} \cdot \nicefrac{H_\mathrm{s}}{2}$ is defined based on the significant wave height $H_\mathrm{s}$. For more details on the HOS simulation, consider the work of \cite{Wedler2023} or \cite{Luenser2022}, for example.  

In this study, we select a wave domain length of $4000 \, \mathrm{m}$, discretized by $n_x=1024$ grid points, resulting in $\Delta x =3.906 \, \mathrm{m}$. A peak enhancement factor of $\gamma=3$ is employed to emulate North Sea conditions. The water depth is $d=500 \, \mathrm{m}$ and the sea state parameters peak wavelength $L_\mathrm{p}$ and steepness $\epsilon$ are varied systematically over $L_\mathrm{p} \in \{80,90, \hdots, 190, 200\}\, \mathrm{m}$ and $ \epsilon \in \{0.01,0.02, \hdots, 0.09, 0.10\} $, resulting in 130 possible $L_\mathrm{p}$-$\epsilon$-combinations. For each $L_\mathrm{p}$-$\epsilon$-combination, we generate four different initial surfaces $\eta(x,t_\mathrm{s}=0)$ by superimposing the wave components of the JONSWAP spectrum with random phase shifts. The subsequent wave evolution $\eta(x, t>0)$ for $t=0, \dots, 50 \, \mathrm{s}$ with $\Delta t_\mathrm{save}=0.1 \, \mathrm{s}$ is performed considering the nonlinearities imposed by HOS order $M=4$. As a result, we generate a total of 520 unique spatio-temporal HOS wave data arrays, each of shape $E_\mathrm{HOS} \in \mathbb{R}^{1024 \times 500}$, where $(E_\mathrm{HOS})_{kj} = \eta (x_k,t_j)$ with  $x_k =  \cdot \Delta x$ and $t_j = j \cdot \Delta t_\mathrm{save}$.

\subsection{Corresponding synthetic radar data}

\begin{figure}[ht]
\vspace{0.1cm}
\centering
\begin{footnotesize}
\includegraphics{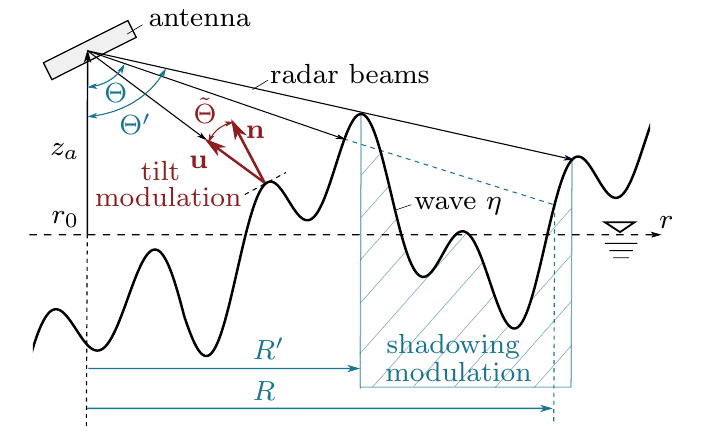}
\end{footnotesize}
\caption{Geometric display of tilt- and shadowing modulation. Tilt modulation $\mathcal{T}(r,t)$ is characterized by the local incidence angle $\Tilde{\Theta}$ between surface normal vector $\mathbf{n}$ and antenna vector $\mathbf{u}$, while shadowing modulation $\mathcal{S}(r,t)$ of a wave facet occurs if another wave closer to the radar systems obstructs the radar beams.}
\label{fig:tilt+shadowing}
\end{figure}

As X-band radar systems are often pre-installed on marine structures for navigation and object detection purposes, they also gained attention for observing ocean surface elevations \citep{Borge1999}. The system antenna rotates with a device-specific revolution time $\Delta t_\mathrm{r}$ of between $1-2\, \mathrm{s}$ \citep{Neill2017} while emitting radar beams along a range $r$. These radar beams interact with short-scale capillary waves distributed on large-scale ocean surface waves by the Bragg resonance phenomenon, resulting in backscatter to the antenna \citep{Valenzuela1978}. This procedure provides measurement data $\xi(r,t)$ as a proxy of wave surface elevations $\eta(r,t)$, which are not directly relatable to each other due to the  influence of different modulation mechanisms. Most influential are assumed to be \textit{tilt modulation} \citep{Dankert2004}, \textit{shadowing modulation}
\citep{NietoBorge2004, Wijaya2015} or a combination of both \citep{Salcedo-Sanz2015}. In order to generate synthetic radar snapshots for this work, the modulation mechanisms are simulated according to \cite{Salcedo-Sanz2015} and \cite{NietoBorge2004}, as illustrated in Figure~\ref{fig:tilt+shadowing}. 

\textit{Tilt modulation} refers to the variation in radar backscatter intensity depending on the local incidence angle $\Tilde{\Theta}(r,t)$ between the unit normal vector $\mathbf{n}(r,t)$ perpendicular to the illuminated wave facet $\eta(r,t)$ and the unit normal vector $\mathbf{u}(r,t)$ pointing towards the antenna. As the backscatter cannot reach the antenna if the dot product $\mathbf{n} \cdot \mathbf{u}$ approaches negative values for $|\Tilde{\Theta}| > \frac{\pi}{2}$, the tilt modulation $\mathcal{T}$ is simulated by
\begin{equation}
    \mathcal{T}(r,t) =
    \mathbf{n}(r,t) \cdot \mathbf{u}(r,t) = \cos \Tilde{\Theta}(r,t)  \hspace{0.5cm}   \text{if} \;\; |\Tilde{\Theta}(r,t)| \le \frac{\pi}{2}
\end{equation}
The \textit{shadowing modulation} instead occurs when high waves located closer to the antenna obstruct waves at greater distances. Shadowing depends on the nominal incidence angle $\Theta(r,t)$ of a wave facet $\eta(r,t)$ with horizontal distance $R(r)$ from the antenna at height $z_\mathrm{a}$ above the mean sea level, geometrically expressed as
\begin{equation}
    \Theta(r,t) = \tan^{-1} \left[ \frac{R(r)}{z_\mathrm{a}-\eta(r,t)} \right].
    \label{eq:shadowing_theta}
\end{equation}
At a specific time instance $t$, a wave facet $\eta(r,t)$ at point $r$ is shadowed in case there is another facet $\eta' = \eta(r',t)$ closer to the radar $R'= R(r')<R(r)$ that satisfies the condition $\Theta' = \Theta(r',t) \ge \Theta(r,t)$. The shadowing-illumination mask $\mathcal{S}$ can be constructed from this condition as follows
\begin{equation}
\mathcal{S}(r,t)=
    \begin{cases}
        0   & \hspace{0.5cm} \text{if} \;\; R(r')<R(r) \text{ and } \Theta(r',t) \ge \Theta(r,t), \\
        1   & \hspace{0.5cm} \text{otherwise}.
    \end{cases}
    \label{eq:shadowing-mask}
\end{equation}
Assuming that tilt-and shadowing modulation contribute to the radar imaging process, the image intensity is proportional to the local radar cross-section, that is $\xi(r,t) \sim \mathcal{T}(r,t)\cdot \mathcal{S}(r,t)$. As marine radars are not calibrated, the received backscatter $\xi(r,t)$ may be normalized to a user-depended range of intensity values.

This work aims to develop a robust ML reconstruction method capable of handling even suboptimal antenna installation conditions. For this reason, we consider a X-band radar system with a comparatively low antenna installation height of $z_\mathrm{a} =18 \, \mathrm{m}$. This choice causes an increased amount of shadowing-affected areas in radar images, which can be inferred from Equations \eqref{eq:shadowing_theta} and \eqref{eq:shadowing-mask}. Around this antenna exists a system's dead range $r_\mathrm{min}$ where the radar beams cannot reach the water surface. In this study, we estimate $r_\mathrm{min}=100 \, \mathrm{m}$,  which is again a comparatively small value and results in the increased magnitude of the tilt modulation influence close to the radar. Moreover, the radar scans the wave surface with a spatial range resolution of $\Delta r = 3.5 \, \mathrm{m}$ at $n_r = 512$ grid points. Thus, the maximum observation range is computed as $r_\mathrm{max}=1892 \, \mathrm{m}$. The radar revolution period is chosen according to \ref{cri:3} to be a snapshot each $\Delta t_\mathrm{r} = 1.3 \, \mathrm{s}$, i.e., $n_t=38$ radar snapshots for $50 \, \mathrm{s}$ of simulation time. Using these definitions, we first transform the 520 wave data arrays $E_\mathrm{HOS} \in \mathbb{R}^{1024 \times 500}$ from their HOS grid to the radar system's grid, yielding $ E_\mathrm{sys} \in \mathbb{R}^{512 \times 38} $, where $(E_\mathrm{sys})_{kj} =\eta (r_k,t_j)$ with  $r_k = i \cdot \Delta r$ and $t_j = j \cdot \Delta t_\mathrm{r}$. To  obtain highly realistic corresponding radar observations, we model tilt modulation $\mathcal{T}(r,t)$ and shadowing modulation $\mathcal{S}(r,t)$, resulting in 520 radar data arrays, each denoted as $ Z_\mathrm{sys} \in \mathbb{R}^{512 \times 38}$ with $(Z_\mathrm{sys})_{kj} =\xi (r_k,t_j)$.

\subsection{Preparation of data for machine learning}
\label{sec:data_preparation}

\begin{figure}[ht]
\centering
\begin{footnotesize}
\includegraphics[width=0.85\textwidth]{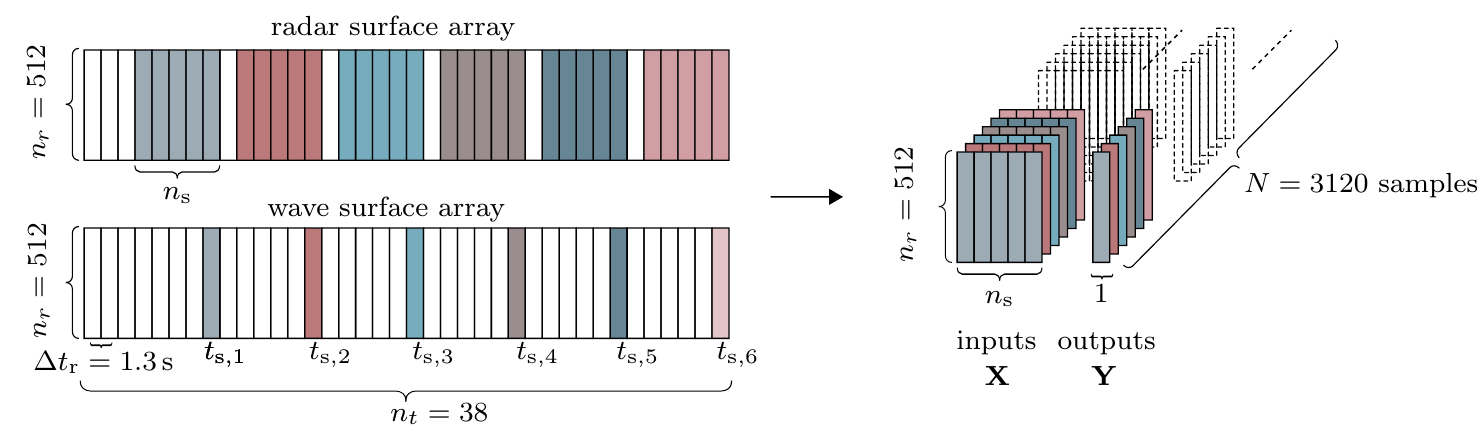}
\end{footnotesize}
\caption{Schematic representation of the ML training sample extraction process. The left-hand side illustrates one of the raw radar and wave surface simulations ($Z_\mathrm{sys}, E_\mathrm{sys} \in \mathbb{R}^{512 \times 38}$), which are utilized to extract input-output samples shown on the right-hand side. Each input $\mathbf{x}_i$ consists $n_\mathrm{s}$ radar snapshots acquired at intervals of $\Delta t_\mathrm{r}=1.3 \, \mathrm{s}$, while each output $\mathbf{y}_i$ represents a single-snapshot wave surface elevation at time instant $t_\mathrm{s}$. In total $N=6 \cdot 520=3120$ data samples are generated.}
\label{fig:dataset}
\end{figure}

To train a supervised learning algorithm, labelled input-output data pairs are required. As visualized in Figure~\ref{fig:dataset}, from each of the 520 generated radar-wave arrays-pairs we extract six radar input snapshots $\mathbf{x}_i$ from the radar surface array $Z_\mathrm{sys}$ and wave output snapshots $\mathbf{y}_i$ from the wave surface array $E_\mathrm{sys}$ at six distinct time instances $t_\mathrm{s}$ with the largest possible temporal distance. Each output sample $\mathbf{y}_i\in \mathbb{R}^{512 \times 1}$ contains a single snapshot at time $t_\mathrm{s}$, while each input sample $\mathbf{x}_i  \in \mathbb{R}^{512 \times n_\mathrm{s}}$ can incorporate a number of $n_\mathrm{s}$ historical radar snapshots at discrete, earlier times $\{ t_\mathrm{s} - j \cdot \Delta t_\mathrm{r} \}_{j=0, \dots, n_\mathrm{s}-1}$. A single snapshot ($n_\mathrm{s}=1$) at a time $t_\mathrm{s}$ can be used as input, however, as we assumed in \ref{hyp:3}, that larger temporal context may enhance the quality of a network's reconstruction $\mathbf{\hat{y}}_i$. Therefore, the optimal value of $n_\mathrm{s}$ is also a subject of investigation as discussed in Sections \ref{sec:UNet_multiple_snap} and \ref{sec:FNO_multiple_snap}. In total, $N=6 \cdot 520 =3120$ input-output data pair samples are generated, each corresponding to a descriptive $L_\mathrm{p}$-$\epsilon$-combination. The data set takes the of shape $\mathbf{X}= [\mathbf{x}_1, \hdots, \mathbf{x}_N]^\mathrm{T} \in \mathbb{R}^{3120\times 512 \times n_\mathrm{s}}$ and $\mathbf{Y}=[\mathbf{y}_1, \hdots, \mathbf{y}_N]^\mathrm{T}  \in \mathbb{R}^{3120 \times 512 \times 1}$ and is split into $60\%$ training, $20\%$  validation, and $20\%$ test data using a stratified data split w.r.t. the sea state parameters $(L_\mathrm{p}, \epsilon)$. This ensures an equal representation of each wave characteristic in the resulting subsets, as described in detail in \ref{sec:Appendix-Hyperparameters}.

\section{Machine learning methodology}
\label{sec:ML Methodolgy}

The U-Net \citep{Ronneberger2015} and the Fourier neural operator (FNO) \citep{Li2020} are neural network architectures for data with grid-like structures such as our radar and wave surface elevation snapshots. Their fundamental difference is the inductive bias encoded by each architecture, which refers to prior assumptions about either the solution space or the underlying data-generating process \citep{Mitchell1980, Battaglia2018}. The U-Net is a special type of CNN \citep{Lecun1989} and imposes an inductive bias by assuming that adjacent data points in Euclidean space are semantically related and learns local mappings between input patches and output features in each layer. This local information is aggregated into more global features due to the utilization of multiple downsampling and convolutional layers. In contrast, the FNO operates under the assumption that the data information can be meaningfully represented in Fourier space. It employs multiple Fourier transformations to learn a mapping between the spectral representation of the input and desired output, directly providing a global understanding of the underlying patterns in the data. This section presents the U-Net- and FNO-based architectures used in our study for radar inversion. In addition, suitable loss and metric functions are introduced for assessing the model's performance.

\subsection{U-Net-based network architecture}
\label{sec:U-Net methodology}

We first adopt the U-Net concept, originally developed for medical image segmentation by \cite{Ronneberger2015}, which has since been applied to a variety of image-to-image translation and surrogate modelling problems, for instance by \cite{Isola16,Liu2018, Stoian2019, Wang2020, Eichinger2020, Niekamp2023} and \cite{Stender2023}. 
The mirrored image dimensions in a fully convolution autoencoder network allow for the U-Net's key property, that is the use of skip-connections for concatenating the output features from the encoding path with the inputs in the decoding path. This enables the reuse of data information of different spatial scales that would otherwise be lost during downsampling and assists the optimizer to find the minimum more efficiently \citep{Li2018}. 

Our proposed encoder-decoder architecture is the result of a four-fold cross-validated hyperparameter study, documented in Table~\ref{tab:hyerparametersUNet} in the appendix. As depicted in Figure~\ref{fig:U-Net}, the adapted U-Net architecture, has a depth of $n_\mathrm{d}=5$ consecutive encoder blocks followed by the same number of consecutive decoder blocks with skip-connections between them.
\begin{figure}[ht]
\vspace{-0.4cm}
\centering
\begin{footnotesize}
\includegraphics[width=0.65\textwidth]{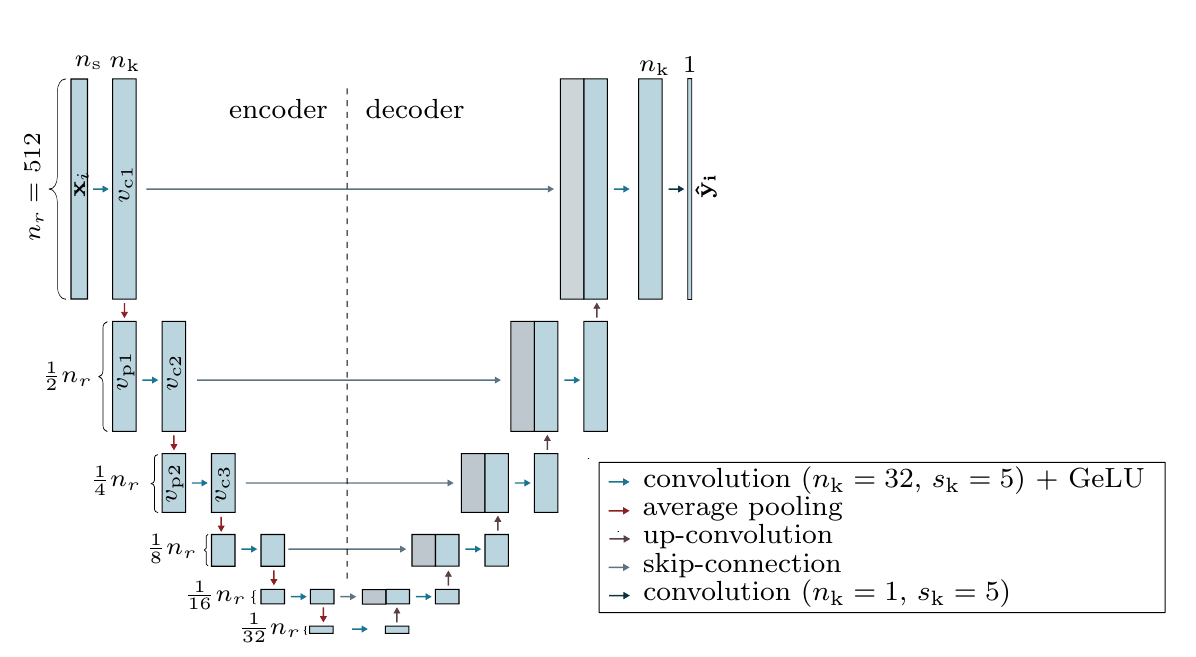}
\end{footnotesize}
\caption{Fully convolutional encoder-decoder architecture based on the U-Net \citep{Ronneberger2015}. Each input $\mathbf{x}_i$ is processed by  $n_\mathrm{d}=5$ alternating convolutional-, activation- and average pooling layers in the encoding path. The decoding path contains convolutional-, activation- and transpose convolutional layers for a gradual upsampling to calculate the output $\mathbf{\hat{y}}_i$. Moreover, the outputs of the encoding stages are transferred to the decoding path via skip-connections.}
\label{fig:U-Net}
\end{figure} 

In more detail, each encoder block in our U-Net-based architecture is composed of a 1D convolutional layer with $n_\mathrm{k}=32$ kernels of size $s_\mathrm{k}=5$, that are responsible for identifying specific features in the input by shifting the smaller-sized kernels, containing the networks trainable weights, across the larger input feature maps in a step-wise manner. Each convolutional layer is followed by a GeLU activation function $\sigma$ \citep{Hendrycks16} and an average pooling downsampling layer of size 2. To summarize, in the encoding path each radar input sample $ \mathbf{x}_i \in \mathbb{R}^{n_r \times n_\mathrm{s}}$ is transformed by the first convolutional layer resulting in $v_\mathrm{c1} \in \mathbb{R}^{n_r \times n_\mathrm{k}}$, with $n_r=512$ being the number of spatial grid-points and $n_\mathrm{s}$ being the historic snapshots in the radar input. Subsequently, this intermediate output is send through $\sigma$, before the pooling layer reduces the spatial dimension to $v_\mathrm{p1} \in \mathbb{R}^{\frac{1}{2} n_r \times n_\mathrm{k}}$. This process is repeated until the final encoding block's output is $v_\mathrm{p5} \in \mathbb{R}^{\frac{1}{32} n_r \times n_\mathrm{k}}$. 
Next, the decoding blocks are applied, each consisting of a convolutional layer with again $n_\mathrm{k}=32$ kernels of size $s_\mathrm{k}=5$, followed by GeLU activation. Afterwards, the feature maps' spatial dimensions are upsampled using transpose convolutional layers with linear activation. The resulting feature maps then are concatenated with the output of the corresponding stage in the encoding path via skip-connections, before the next convolution is applied. This process is repeated until the final wave output $\mathbf{\hat{y}}_i \in \mathbb{R}^{n_r \times 1}$ is calculated using a convolutional layer with a single kernel and linear activation.

As indicated above, the U-Net architecture assumes local connections between neighbouring data points, which is accomplished through two mechanisms. Firstly, the convolutional layers use kernels with a receptive field of $s_\mathrm{k}=5$ pixels to process different local parts of the larger input feature maps in the same manner. This is referred to as weight sharing, causing a property called \textit{translational equivariance}: each patch of the input is processed by the same kernels. Secondly, the pooling layers induce locality by assuming that meaningful summations of information from small local regions in the intermediate feature maps can be made and creates a property referred to as \textit{translational invariance} \citep{Goodfellow2016}.

\subsection{FNO-based network architecture}
\label{sec:FNO methodology}

In the second step, we explore a neural network based on the FNO \citep{Li2020}. While a CNN is limited to map between finite-dimensional spaces, neural operators are in addition capable to learn nonlinear mappings between a more general class of function spaces. This makes the FNO well-suited for capturing the spatio-temporal patterns that govern the dynamics of various physical problems that obey partial differential equations if the solutions are well represented in Fourier space. FNO variants have been applied to e.g., fluid dynamics \citep{Peng2022, Li2022}, simulation of multiphase flow  \citep{Yan2022, Wen2022}, weather forecasting \citep{Pathak2022}, material modeling \citep{Rashid2022, You2022}, and image classification \citep{Williamson2022}.

The FNO-based iterative architecture approach ($\mathbf{x}_i \rightarrow v_0 \rightarrow v_1 \rightarrow \hdots \rightarrow  \mathbf{\hat{y}}_i$) applied in this work is illustrated in Figure~\ref{fig:FNO}, while Table~\ref{tab:hyperparametersFNO} in the appendix summarizes the determination of model hyperparameters by four-fold cross-validation. 

\begin{figure}[ht]
\centering
\begin{footnotesize}
\includegraphics[width=0.95\textwidth]{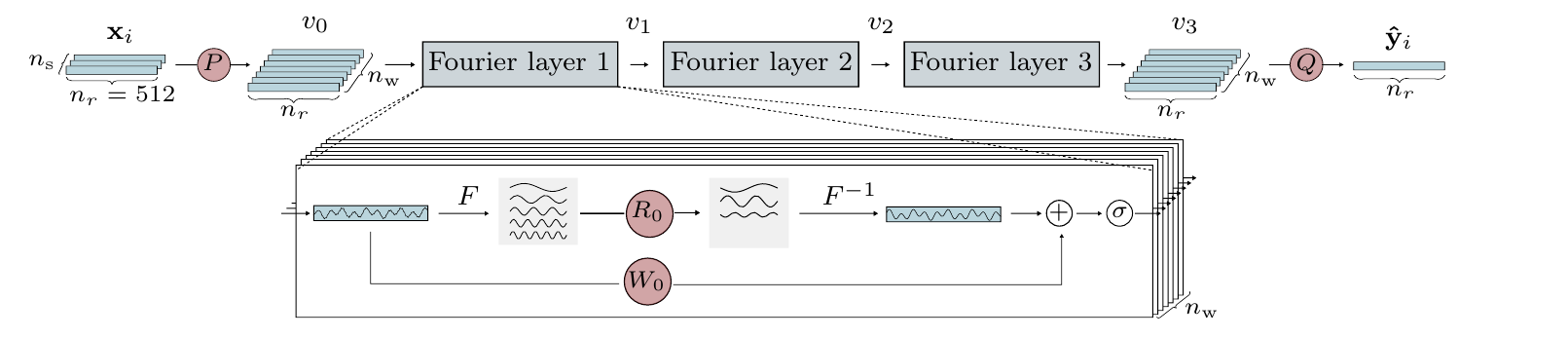}
\end{footnotesize}
\caption{Network architecture based on the Fourier neural operator \citep{Li2020}. Each input $\mathbf{x}_i$ is lifted to a higher dimensional representation $v_0$  of channel width $d_\mathrm{w}$ by a neural network $P$. Afterwards, $n_\mathrm{f}=3$ Fourier layers are applied to each channel. Finally, $v_3$ is transferred back to the target dimension of the output $\mathbf{\hat{y}}_i$ by another neural network $Q$. More specifically, each Fourier layer is composed of two paths. The upper one learns a mapping in Fourier space by adapting $R_j$ for scaling and truncating the Fourier Series after $n_\mathrm{m}$ modes, while the lower one learns a local linear transform $W_j$.}
\label{fig:FNO}
\end{figure}

The proposed FNO transforms radar input data $\mathbf{x}_i \in \mathbb{R}^{n_r \times n_\mathrm{s}}$ into a higher-dimensional latent representation $v_0 \in \mathbb{R}^{n_r \times n_\mathrm{w} }$ of channel width $n_\mathrm{w} = 32$, using a linear neural network layer $P$ with $n_\mathrm{w}$ nodes. Subsequently, the latent representation passes through $n_\mathrm{f}=3$ Fourier layers, each consisting of two paths. In the upper path, a global convolution operator defined in Fourier space is applied to each channel of $v_0$ separately utilizing discrete Fourier transforms $F$. A linear transformation $R_0$ is then applied to the lower-order Fourier modes after truncating the Fourier series at a maximum number of $n_\mathrm{m}=64$ modes. Subsequently, this scaled and filtered content is back-transformed to the spatial domain using inverse discrete Fourier transforms $F^{-1}$. In the lower path, a linear transformation $W_0$ in the spatial domain is applied to the input $v_0$ to account for non-periodic boundary conditions and higher-order modes that are neglected in the upper path of the Fourier layer. The outputs of the upper and lower paths are added, and the sum is passed through a nonlinear GeLU activation $\sigma$ resulting in $v_1 \in \mathbb{R}^{n_r \times n_\mathrm{w} }$, before entering the next Fourier layer. In summary, the output of the $(j+1)$-th Fourier layer is defined as
\begin{equation}
    v_{j+1} = \sigma \left( F^{-1} \left(R_j \cdot F(v_j) \right) + W_j \cdot v_j \right).
\end{equation}
Finally, the output $v_3$ of the last Fourier layer is transferred to the target wave output dimension $\mathbf{\hat{y}}_i \in \mathbb{R}^{ n_r \times 1}$ using another linear layer $Q$. In summary, the FNOs weights correspond to $P \in \mathbb{R}^{n_\mathrm{s} \times d_\mathrm{w}}$, $Q \in \mathbb{R}^{n_\mathrm{s} \times d_\mathrm{w}}$ and all $R_j \in \mathbb{C}^{d_\mathrm{w} \times d_\mathrm{w} \times d_\mathrm{m}}$ and $W_j \in \mathbb{R}^{d_\mathrm{w} \times d_\mathrm{w}}$. As the $R_j$-matrices contain the main portion of the total number of  weights, most parameters are learned in the Fourier space rather than the original data space.

As previously noted, the FNO architecture incorporates a global inductive bias that assumes the input data exhibits approximately periodic properties and can be effectively represented in Fourier space. Furthermore, the FNO's design presupposes that the Fourier spectrum of the input data is smooth, enabling its frequency components to be represented by a limited number of low-wavenumber Fourier coefficients, as the $R_j$ matrices, which are responsible for the global mapping, truncate higher-frequency modes. 

\subsection{Training and evaluation}
\label{sec:Training+evaluation}

Both the U-Net- and FNO-based architecture are implemented using the PyTorch library \citep{Pytorch2019}. To enable a fair comparison and account for wave training data of varying spatial scales, the mean of the relative L2-norm of the error is employed as loss function $\mathcal{L}$ for both architectures. The relative L2-norm error for one sample $i$ is defined as follows, where $\mathbf{y}_i$ and $\mathbf{\hat{y}}_i \in \mathbb{R}^{512 \times 1}$ represent the true and reconstructed wave surface
\begin{equation}
     \mathrm{nL2}(\mathbf{y}_i, \mathbf{\hat{y}}_i) = \mathrm{nL2}_i = \frac{\| \mathbf{\hat{y}}_i - \mathbf{y}_i \|_2}{\| \mathbf{y}_i \|_2 }.
\end{equation}
While we use the subscript $i$ to represents a sample-specific error $\mathrm{nL2}_i$, the value $\mathrm{nL2}$ without a subscript denotes the mean value across a number of samples $N$, for example the mean error across the training set $\mathcal{L}:= \mathrm{nL2} = \frac{1}{N_\mathrm{train}} \sum_{i=1}^{N_\mathrm{train}} \mathrm{nL2}(\mathbf{y}_i, \mathbf{\hat{y}}_i)$. To minimize the loss, we use the Adam optimizer \citep{Kingma2014} with a learning rate of 0.001. The training is executed for 800 epochs on an NVIDIA GeForce RTX 3050 Ti Laptop GPU. For both, the U-Net-based models $\mathcal{M}_{\mathrm{U}, n_\mathrm{s}}$ and FNO-based models $\mathcal{M}_{\mathrm{F}, n_\mathrm{s}}$, only the models with the lowest test loss within the 800 epochs is stored for performance evaluation and visualization.

Established machine learning metrics based on Euclidean distances treat the deviation of two surfaces in frequency or phase as amplitude errors \citep{WEDLER2022}. Therefore, we introduce the surface similarity parameter (SSP) proposed by \cite{Perlin2014} as an additional performance metric
\begin{equation}
    \mathrm{SSP}(\mathbf{y}_i, \mathbf{\hat{y}_i}) = \mathrm{SSP}_i = \frac{\sqrt{\int | F_{\mathbf{y}_i}(k)-F_{\mathbf{\hat{y}}_i}(k)|^2 dk}}{\sqrt{\int | F_{\mathbf{y}_i}(k)|^2 dk}+\sqrt{\int |F_{\mathbf{\hat{y}}_i}(k)|^2 dk}} \in [0, 1],
    \label{eq:SSP}
\end{equation}
where $k$ denotes the wavenumber vector and $F_{\mathrm{y}_i}$ denotes the discrete Fourier transform of a surface $\mathbf{y}_i$. The SSP is a normalized error metric, with $\mathrm{SSP}_i=0$ indicating perfect agreement and $\mathrm{SSP}_i=1$ a comparison against zero or of phase-inverted surfaces. As the SSP combines phase-, amplitude-, and frequency errors in a single quantity, it is used in recent ocean wave prediction and reconstruction studies by \cite{Klein2020, Klein2022}, \cite{WEDLER2022, Wedler2023}, \cite{Desmars2021, Desmars2022} and \cite{Luenser2022}. 

While metrics such as the $\mathrm{nL2}_i$ or $\mathrm{SSP}_i$ evaluate the average reconstruction quality of each $\mathbf{\hat{y}}_i \in \mathbb{R}^{n_r \times 1}$ across the entire spatial domain $r$ with $n_r=512$ grid points, it is important to consider the potential imbalance in reconstruction error between those areas where the radar input $\mathbf{x}_i$ was either shadowed or visible. This imbalance ratio can be quantified by $\tfrac{\mathrm{nL2}_{\mathrm{shad}_i}}{\mathrm{nL2}_{\mathrm{vis}_i}}$. Here, $\mathrm{nL2}_{\mathrm{shad}_i} = \mathrm{nL2}(\mathbf{y}_{\mathrm{shad}_i}, \mathbf{\hat{y}}_{\mathrm{shad}_i})$ and $\mathrm{nL2}_{\mathrm{vis}_i} = \mathrm{nL2}(\mathbf{y}_{\mathrm{vis}_i}, \mathbf{\hat{y}}_{\mathrm{vis}_i})$ are the errors of the output wave elevations in the shadowed or visible areas, respectively. We separate the visible and shadowed parts using the shadowing mask $\mathcal{S}$ introduced in Eq.~\eqref{eq:shadowing-mask}, where  $\mathbf{y}_{\mathrm{vis}_i}= \mathcal{S} \cdot \mathbf{y}_i$ and $\mathbf{y}_{\mathrm{shad}_i}= (1-\mathcal{S}) \cdot \mathbf{y}_i$. Afterwards, all cells with zero entries are removed from the output arrays, such that the number of visible or invisible data points is $n_{\mathrm{vis}_i}$ or $n_{\mathrm{shad}_i}$, respectively, and $\mathbf{y}_{\mathrm{vis}_i}, \, \mathbf{\hat{y}}_{\mathrm{vis}_i}  \in \mathbb{R}^{n_{\mathrm{vis}_i} \times 1}$ and $\mathbf{y}_{\mathrm{shad}_i}, \, \mathbf{\hat{y}}_{\mathrm{shad}_i} \in \mathbb{R}^{n_{\mathrm{shad}_i} \times 1}$ satisfy $n_{\mathrm{vis}_i}+n_{\mathrm{shad}_i}=n_r=512$. To conclude, a high value of the ratio indicates that the reconstruction in areas that were shadowed in the input is much worse than in the visible areas. We thus not only strive for low $\mathrm{nL2}_i$ values, but also for low $\tfrac{\mathrm{nL2}_{\mathrm{shad}_i}}{\mathrm{nL2}_{\mathrm{vis}_i}}$ ratios to achieve uniform reconstructions. We use a ratio metric only based on the Euclidean distance based $\mathrm{nL2}_i$ and not for the $\mathrm{SSP}_i$, as small sections of $\mathbf{y}_i$ and $\mathbf{\hat{y}}_i$ cannot be meaningfully considered in Fourier space.

%%% Results %%%
\section{Results}
\label{sec:results}

This work explores the potential of utilizing machine learning for the reconstruction of one-dimensional ocean wave surfaces $\eta$ from radar measurement surfaces $\xi$ at a time instance $t_\mathrm{s}$. Therefore, each radar input sample $\mathbf{x}_i \in \mathbb{R}^{n_r \times n_\mathrm{s}}$, with $n_r=512$ being the number of spatial grid points in range direction and $n_\mathrm{s}$ being the number of radar snapshots, is acquired according to Section~\ref{sec:data generation}. Each input $\mathbf{x}_i$ is to be mapped to the desired wave surface output $\mathbf{y} _i\in \mathbb{R}^{n_r \times 1}$ via a ML model $\mathcal{M}$. We examine the impact of the inductive bias of the U-Net-based models $\mathcal{M}_{\mathrm{U}, n_\mathrm{s}}$ and the FNO-based models $\mathcal{M}_{\mathrm{F}, n_\mathrm{s}}$ proposed in Section~\ref{sec:ML Methodolgy}, as well as the impact of the number of historical radar snapshots $n_\mathrm{s}$ included in each input $\mathbf{x}_i$. We train the models using a total data set of $N_\mathrm{train}=2496$ samples and thus to learn the mapping $\mathcal{M}:\mathbf{X}\rightarrow \mathbf{Y}$ with $ \mathbf{X}\in \mathbb{R} ^{2496 \times n_r \times n_\mathrm{s}}, \mathbf{Y}\in \mathbb{R} ^{2496 \times n_r \times 1} $. Afterwards, we evaluate their performance using the previously excluded test set of $N_\mathrm{test}=624$ samples. The results are summarized in Table~\ref{tab:results}, and are discussed regarding the pre-stated \ref{hyp:1}-\ref{hyp:3} and \ref{cri:1}-\ref{cri:3} in detail in the subsequent subsections.

\begin{table}[!ht]
\centering
\caption{Reconstruction results averaged across the entire test set evaluated with different metrics for the U-Net-based models $\mathcal{M}_{\mathrm{U},n_\mathrm{s}}$ and FNO-based models  $\mathcal{M}_{\mathrm{F},n_\mathrm{s}}$ trained with either one or multiple radar snapshots $n_\mathrm{s}$ in each sample's input.}
\begin{scriptsize}
\begin{tabular}{llrccccc}
\toprule
\toprule
\multicolumn{5}{c}{} & \multicolumn{3}{c}{mean errors across }  \\
\multicolumn{5}{c}{model} & \multicolumn{3}{c}{$N_\mathrm{test}=624$ test set samples}   \\
\cmidrule(rl){1-5} \cmidrule(rl){6-8} 
name & architecture & $n_\mathrm{s}$ & epochs & investigated in & $\mathrm{nL2}$ & $\frac{\mathrm{nL2}_\mathrm{shad}}{\mathrm{nL2}_\mathrm{vis}}$ & $\mathrm{SSP}$  \\ 
\midrule
\midrule
$\mathcal{M}_{\mathrm{U},1}$    & U-Net-based   & 1     & 150  & Sec. \ref{sec:UNet_single_snap}    & 0.329 &   2.679   &   0.171 \\
$\mathcal{M}_{\mathrm{U},10}$   & U-Net-based   & 10    & 592  & Sec. \ref{sec:UNet_multiple_snap}  & 0.123 &   1.755   &   0.061\\
$\mathcal{M}_{\mathrm{F},1}$    & FNO-based     & 1     & 721  & Sec. \ref{sec:FNO_single_snap}     & 0.242 &   1.886   & 0.123\\
$\mathcal{M}_{\mathrm{F},9}$    & FNO-based     & 9     & 776  & Sec. \ref{sec:FNO_multiple_snap}   & 0.153 &   1.381   &  0.077\\
\bottomrule
\bottomrule
\end{tabular}
\end{scriptsize}
\label{tab:results}
\vspace{-0.2cm}
\end{table}

\subsection{Performance of the U-Net-based model}

In the first step of our investigation, we examine the ability of U-Net-based models $\mathcal{M}_{\mathrm{U},n_\mathrm{s}}$ to reconstruct wave surfaces along the full spatial dimension, which covers $r_\mathrm{max}-r_\mathrm{min}=1792 \, \mathrm{m}$ on $n_r=512$ grid points. We use the $N_\mathrm{train}=2496$ samples of single snapshot ($n_\mathrm{s}=1$) radar input data for training. Afterwards, we utilize the same architecture to determine the best number of historical snapshots $n_\mathrm{s}>1$ required in the radar inputs to achieve the best reconstruction performance. We also visually compare reconstructed wave elevations $\mathbf{\hat{y}}_i$ of two selected samples from the test set with their corresponding true elevations~$\mathbf{y}_i$.

\subsubsection{U-Net using single-snapshot radar data}
\label{sec:UNet_single_snap}

Mapping of single snapshot radar data ($n_\mathrm{s}=1$) refers to mapping a radar snapshot $\mathbf{x}_i \in \mathbb{R}^{n_r \times 1}$ to a wave snapshot $\mathbf{y}_i \in \mathbb{R}^{n_r \times 1}$, with $n_r=512$ spatial grid points, that are recorded at the same time instant $t_\mathrm{s}$. According to Table~\ref{tab:results} the U-Net-based model $\mathcal{M}_{\mathrm{U},1}$ trained with the available $N_\mathrm{train}=2496$ samples achieves a reconstruction performance given by a mean loss value of $\mathrm{nL2}=0.329$ across all $N_\mathrm{test}=624$ test set samples after 150 epochs of training. Afterwards, the model tends to overfit the training data, as shown in the loss curve in Figure~\ref{fig:loss_UNet_1_snap}. The observed error corresponds to a mean value of $\mathrm{SSP}=0.171$ across all test set samples, which fails to satisfy the \ref{cri:2} of reconstruction errors below $\mathrm{SSP \le 0.1}$.

\begin{figure}[ht!]
     \centering
     \begin{subfigure}[b]{0.49\textwidth}
         \centering
         \includegraphics[width=\textwidth]
         {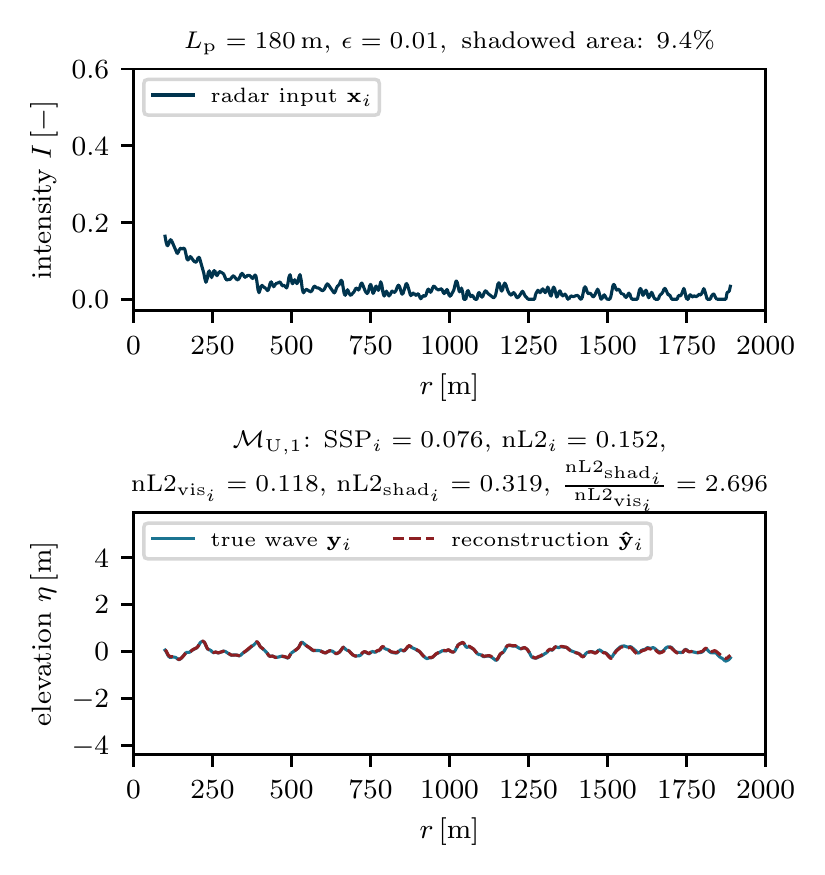}
         \caption{}
         \label{subfig:UNet_1_snap_left}
     \end{subfigure}
     \hfill
     \begin{subfigure}[b]{0.49\textwidth}
         \centering
         \includegraphics[width=\textwidth]{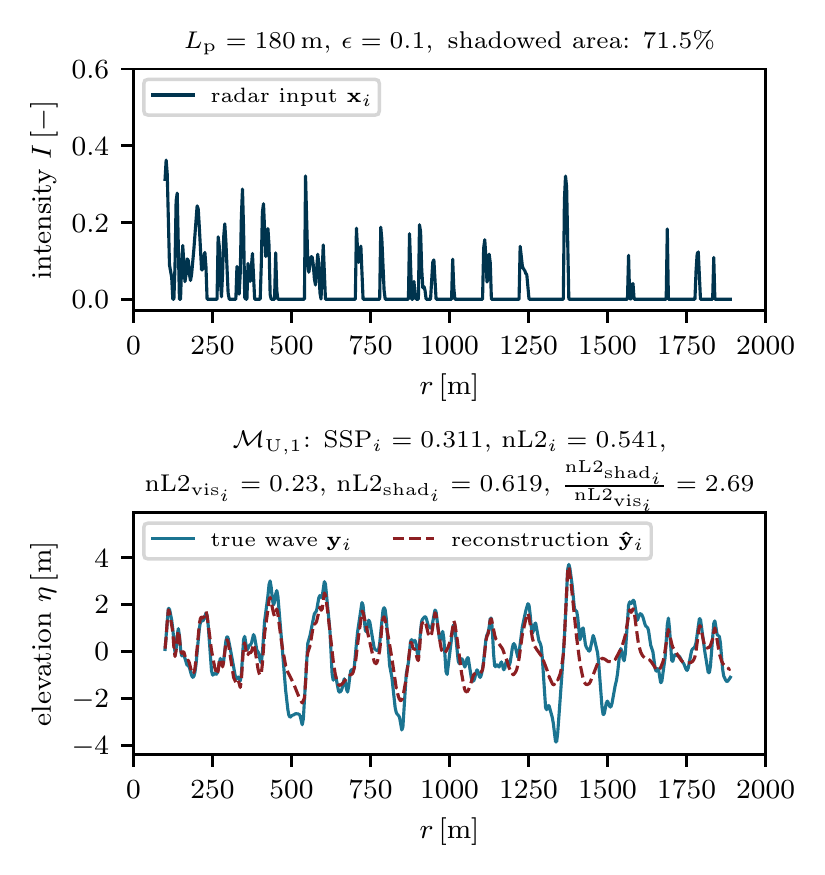}
         \caption{}
         \label{subfig:UNet_1_snap_right}
     \end{subfigure}
     \caption{Two samples from the test set described by the same wavelength $L_\mathrm{p}=180\, \mathrm{m}$, but different steepness $\epsilon$, reconstructed by the U-Net-based architecture $\mathcal{M}_{\mathrm{U},1}$. (a) Small $\epsilon$ values cause minor impact of the shadowing modulation in the radar input and allow accurate reconstructions. (b) Larger $\epsilon$ create more extensive shadowed areas and cause higher reconstruction errors.}
     \label{fig:UNet_1_snap}
\end{figure}

To identify the origin of reconstruction errors, we employed model $\mathcal{M}_{\mathrm{U},1}$ to generate reconstructions~$\mathbf{\hat{y}}_i$ for two exemplary radar input samples $\mathbf{x}_i$ from the test set. Despite the stratified data split ensuring an equal distribution of sea state parameter combinations $(L_\mathrm{p}, \epsilon)$ in the training and test set, the errors are unevenly distributed across individual samples $i$, as exemplarily illustrated in Figure~\ref{fig:UNet_1_snap}: The sample in Figure~\ref{subfig:UNet_1_snap_left} corresponds to a peak wavelength $L_\mathrm{p}=180 \, \mathrm{m}$ and small amplitudes caused by a small steepness of $\epsilon=0.01$. It exhibits a minor impact from the shadowing modulation mechanism only affecting $9.4 \%$ of the total radar-illuminated surface $\mathbf{x}_i$ in the top panel. The corresponding surface reconstruction $\mathbf{\hat{y}}_i$ generated by $\mathcal{M}_{\mathrm{U},1}$ in the bottom panel closely approximates the true wave elevation $\mathbf{y}_i$, as evidenced by the sample-specific error of $\mathrm{nL2}_i=0.152$ or $\mathrm{SSP}_i=0.076$.
In contrast, the second sample in Figure~\ref{subfig:UNet_1_snap_right} with the same $L_\mathrm{p}=180 \, \mathrm{m}$ but increased $\epsilon=0.10$ shows $71.5\%$ of the spatial $r$-domain being affected from shadowing modulation causing zero-valued intensities. This results in a high reconstruction error of $\mathrm{nL2}_i=0.541$ or $\mathrm{SSP}_i=0.311$. Particularly the shadowed areas seem to contribute to the poor reconstruction, as their error is 2.69 times higher than in the visible areas, indicated by $\frac{\mathrm{nL2}_{\mathrm{shad}_i}}{\mathrm{nL2}_{\mathrm{vis}_i}}$.

\subsubsection{U-Net using spatio-temporal radar data}
\label{sec:UNet_multiple_snap}

\begin{figure}[ht!]
\centering
\begin{footnotesize}
\includegraphics{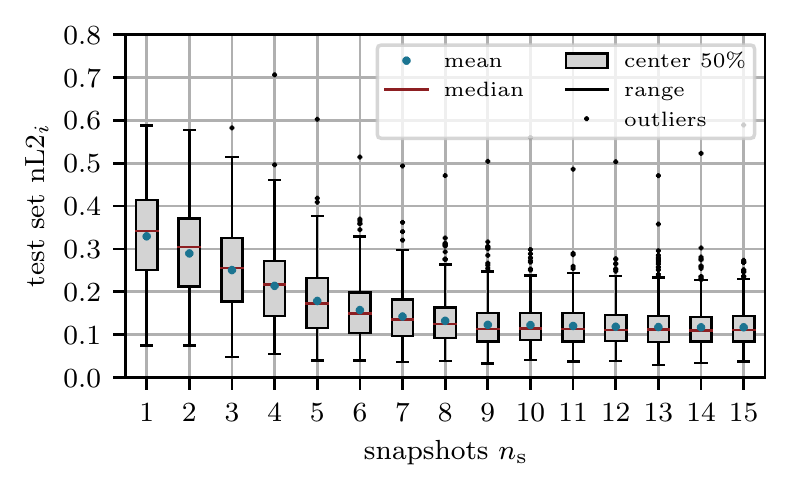}
\end{footnotesize}
\vspace{-0.2cm}
\caption{Boxplot depicting the error distribution on test set, depending on the number of historical radar snapshots $n_\mathrm{s}$ provided to train U-Net-based architectures $\mathcal{M}_{\mathrm{U},n_\mathrm{s}}$. The best model performance is achieved for $n_\mathrm{s} = 10$.} 
\label{fig:boxplotUNet}
\end{figure}

To improve the reconstruction quality of the U-Net-based architecture, especially for high wave steepness, we took inspiration from classical spectral-analysis- and optimization-based reconstruction approaches \citep[cf.][]{NietoBorge2004, Wu2004}. These approaches use spatio-temporal radar data by considering temporal sequences of $n_\mathrm{s}$ historical radar snapshots for reconstruction. Thus, we use multiple historical radar snapshots $n_\mathrm{s}$ that satisfy \ref{cri:3} with $\Delta t_\mathrm{r}=1.3 \, \mathrm{s}$ for each input sample $\mathbf{x}_i \in \mathbb{R}^{512 \times n_\mathrm{s}}$, while the outputs remain single snapshots $\mathbf{y}_i \in \mathbb{R}^{n_r \times 1}$ at the respective last time instant $t_\mathrm{s}$.  We conducted 14 additional training runs of the same architecture using all $N_\mathrm{train}=2496$ input-output samples of the training set, but with increasing $n_\mathrm{s}$ in the inputs $\mathbf{x}_i$, to determine the best number of snapshots $n_\mathrm{s}$. This procedure is summarized in the boxplot in Figure~\ref{fig:boxplotUNet}. 

The boxplot shows that the model's mean performance across the entire test set significantly improves up to a value of $n_\mathrm{s}=10$, confirming \ref{hyp:3} as the reconstruction quality improves by incorporating multiple radar snapshots in the input. Moreover, the sample-specific error values $\mathrm{nL2}_i$ become less scattered around the mean value.
The model $\mathcal{M}_{\mathrm{U},10}$, determined by the boxplot analysis, achieves a final mean reconstruction performance of $\mathrm{nL2}=0.123$ or $\mathrm{SSP}=0.061$ across the $N_\mathrm{test}=624$ test set samples, as shown in Table~\ref{tab:results}, now satisfying \ref{cri:2} of $\mathrm{SSP}\le 0.10$ and thus confirms \ref{hyp:1}. In addition, it yields a lower ratio of $\tfrac{\mathrm{nL2}_\mathrm{shad}}{\mathrm{nL2}_\mathrm{vis}}=1.755$ compared to $2.679$ for model $\mathcal{M}_{\mathrm{U},1}$, indicating a more balanced reconstruction between shadowed and visible areas on average. Moreover, the model does not exhibit early overfitting anymore, achieving the best performance after 592 epochs, shown in Figure~\ref{fig:loss_UNet_10_snap}.

\begin{figure}[ht]
     \centering
     \begin{subfigure}[b]{0.49\textwidth}
         \centering
         \includegraphics[width=\textwidth]
         {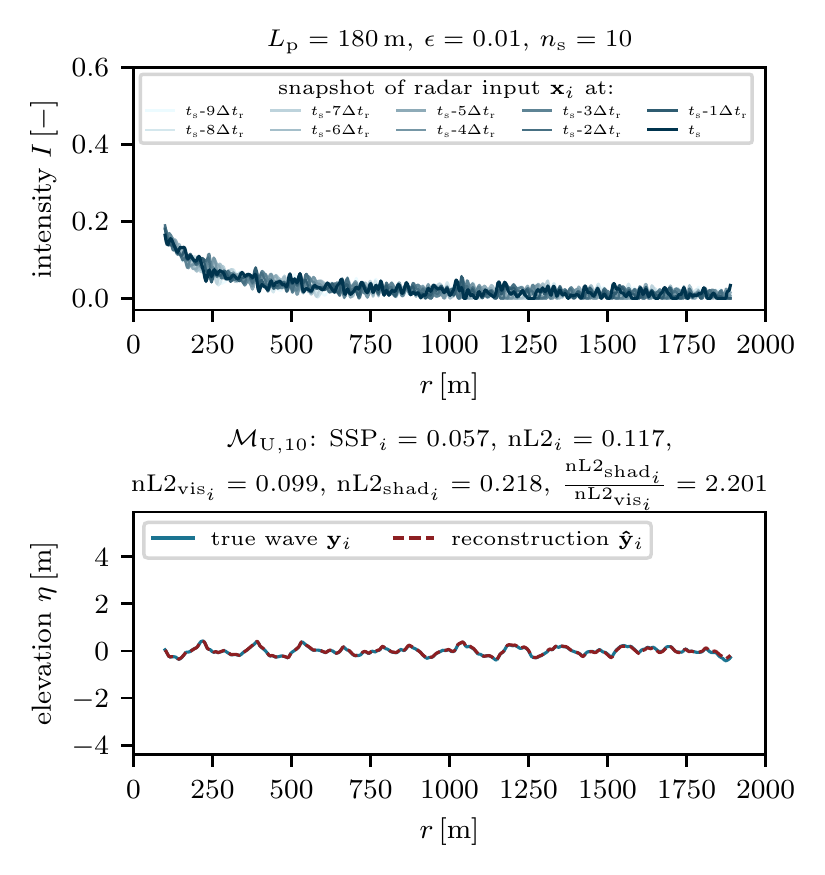}
         \caption{}
         \label{subfig:UNet_10_snaps_left}
     \end{subfigure}
     \hfill
     \begin{subfigure}[b]{0.49\textwidth}
         \centering
         \includegraphics[width=\textwidth]{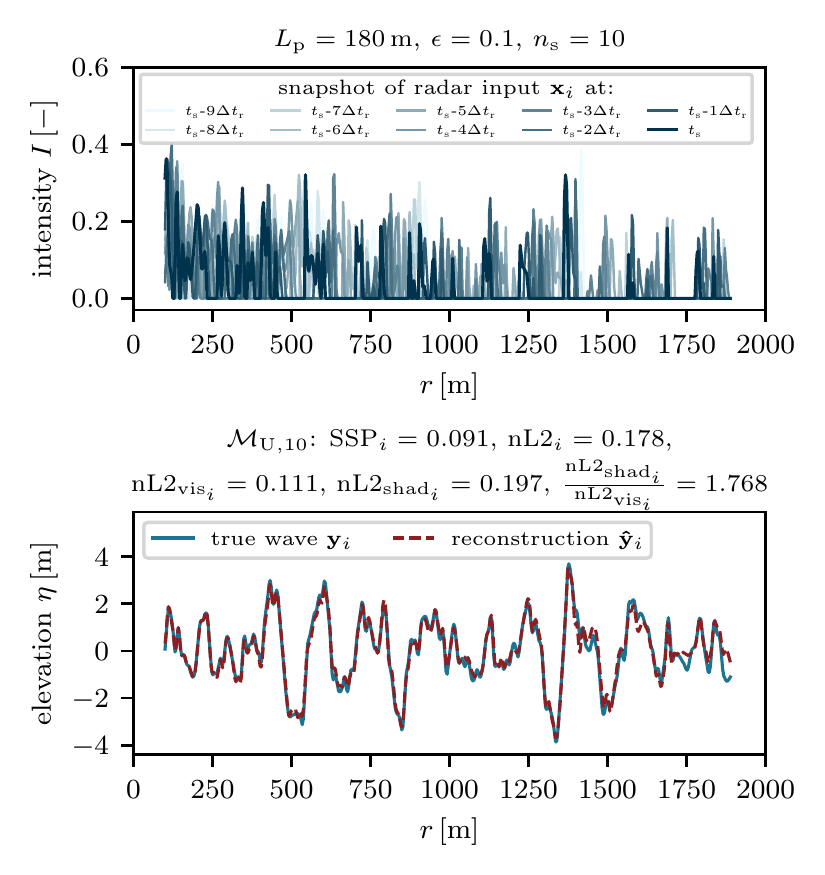}
         \caption{}
         \label{subfig:UNet_10_snaps_right}
     \end{subfigure}
     \caption{Two samples from the test set described by same wavelength $L_\mathrm{p}=180 \, \mathrm{m}$, but different wave steepness $\epsilon$, reconstructed by the U-Net-based architecture trained with $n_\mathrm{s}=10$ historical snapshots in the radar input $\mathcal{M}_{\mathrm{U},10}$. Compared to $\mathcal{M}_{\mathrm{U},1}$, a strong reconstruction improvement is observed, especially for the sample with high $\epsilon=0.10$ in (b).}
     \label{fig:UNet_10_snap}
\end{figure}

Figure~\ref{fig:UNet_10_snap} further confirms the improvement of the reconstruction, using $n_\mathrm{s}=10$ radar snapshots in each input to train $\mathcal{M}_{\mathrm{U},10}$, by depicting the same two exemplary test set samples reconstructed by $\mathcal{M}_{\mathrm{U},1}$ in Figure~\ref{fig:UNet_1_snap} before. The top panels display the most recent ($t_\mathrm{s}$) radar snapshot present in $\mathbf{x}_i \in \mathbb{R}^{n_r \times n_\mathrm{s}}$ in the darkest shading and preceding snapshots at $t_j = \{ t_\mathrm{s}-j \Delta t_\mathrm{r} \}_{j=0, \dots, n_\mathrm{s}-1} $ in increasingly lighter shades. Compared to Figure~\ref{fig:UNet_1_snap}, the sample with small $\epsilon=0.01$ in Figure~\ref{subfig:UNet_10_snaps_left} experiences only a slight reduction in reconstruction error, while the sample with $\epsilon=0.10$ in Figure~\ref{subfig:UNet_10_snaps_right} exhibits a substantial reduction around one-third of the previous sample-specific $\mathrm{nL2}_i$ or $\mathrm{SSP}_i$ value. The improved performance seems mainly attributable to the enhanced reconstruction of shadowed areas.

\subsection{Performance of the FNO-based model}

The U-Net-based model $\mathcal{M}_{\mathrm{U}, 10}$ already supported \ref{hyp:1} and \ref{hyp:3} by demonstrating the potential to reconstruct wave surface elevations from radar data in general and improving the reconstruction quality by including additional historical radar data in the input. However, we also hypothesized that the FNO-based architecture may outperform CNN-based methods, such as the U-Net, due to its global inductive bias (\ref{hyp:2}), which may be beneficial for the wave data structure. To investigate this, we  again use the entire set of $N_\mathrm{train}=2496$ samples to train FNO-based models $\mathcal{M}_{\mathrm{F},{n_\mathrm{s}}}$ with $n_\mathrm{s}=1$ radar snapshot in each of the inputs $\mathbf{x}_i \in \mathbb{R}^{n_r \times n_\mathrm{s}}$ first. Subsequently, we determine the number $n_\mathrm{s}>1$ to achieve the best reconstruction performance. Both investigations again are conducted on the entire domain of $1792 \, \mathrm{m}$ $(n_r=512)$ and we compare true and reconstructed elevations $\mathbf{y}_i$ and $\mathbf{\hat{y}}_i \in \mathbb{R}^{n_r \times 1}$ of two exemplary samples.

\subsubsection{FNO using single-snapshot radar data}
\label{sec:FNO_single_snap}

The FNO-based model $\mathcal{M}_{\mathrm{F},1}$ trained with $n_\mathrm{s}=1$ snapshot in each input, attains its best performance $\mathrm{nL2}=0.240$ after 721 training epochs, as  shown Table \ref{tab:results} and demonstrated in the loss curve in Figure~\ref{fig:loss_FNO_1_snap}. Although the corresponding mean $\mathrm{SSP}=0.123$ across all $N_\mathrm{test}=624$ samples in the test set does not attain the \ref{cri:2}, the error still presents a notable improvement compared to the $\mathrm{SSP}$ value of 0.171 previously obtained by the U-Net-based model $\mathcal{M}_{\mathrm{U},1}$. Moreover, $\mathcal{M}_{\mathrm{F},1}$ not only reduces the mean nL2 or SSP error but also reconstructs the waves more uniformly between shadowed and visible areas compared to $\mathcal{M}_{\mathrm{U},1}$. This is evident by the decrease in the mean $\tfrac{\mathrm{nL2}\mathrm{shad}}{\mathrm{nL2}\mathrm{vis}}$-ratio from $2.679$ to $1.886$.

\begin{figure}[ht]
     \centering
     \begin{subfigure}[b]{0.49\textwidth}
         \centering
         \includegraphics[width=\textwidth]
         {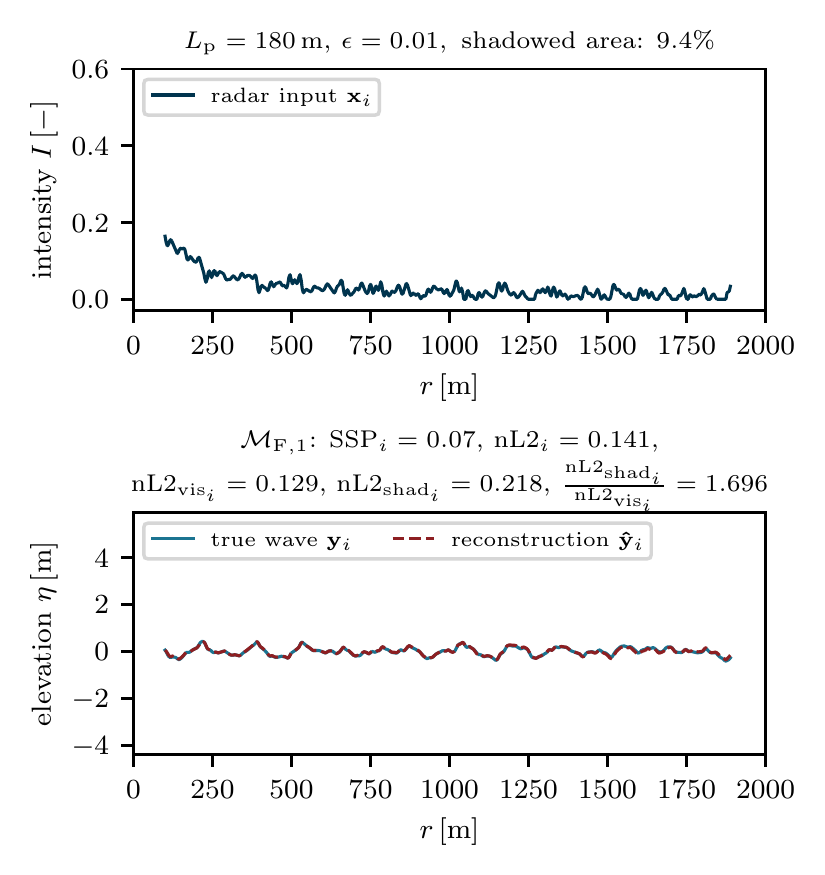}
         \caption{}
         \label{subfig:FNO_1_snap_left}
     \end{subfigure}
     \hfill
     \begin{subfigure}[b]{0.49\textwidth}
         \centering
         \includegraphics[width=\textwidth]{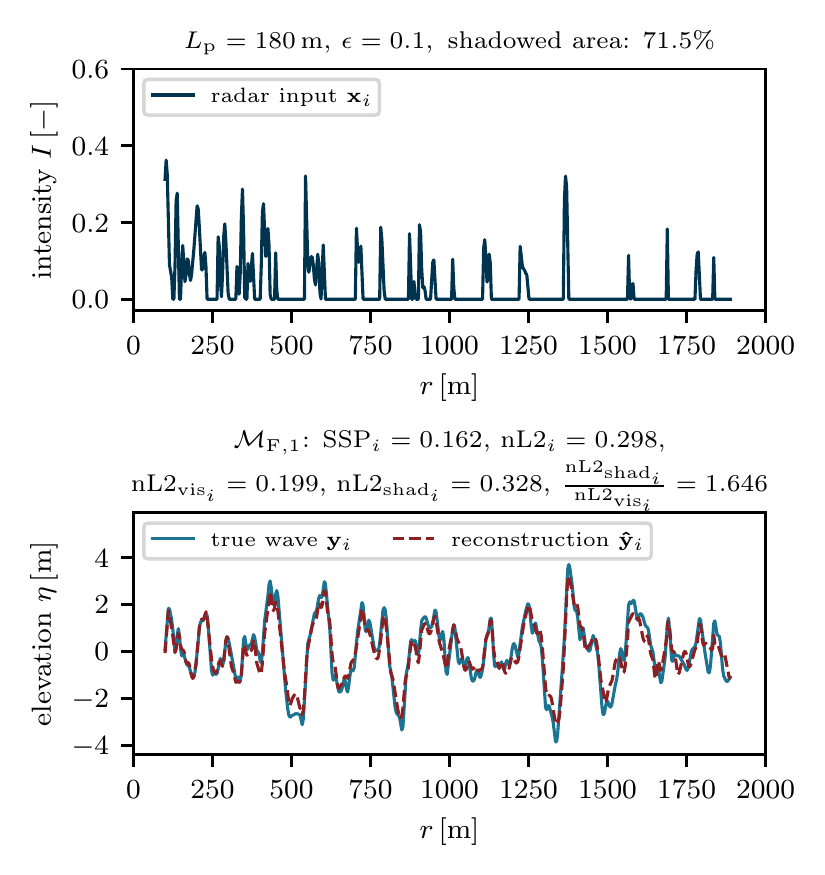}
         \caption{}
         \label{subfig:FNO_1_snap_right}
     \end{subfigure}
     \caption{Two samples from the test set described by the same wavelength $L_\mathrm{p}=180 \, \mathrm{m}$, but different wave steepness $\epsilon$ reconstructed by the FNO-based architecture $\mathcal{M}_{\mathrm{F},1}$. The $\mathcal{M}_{\mathrm{F},1}$ outperforms the $\mathcal{M}_{\mathrm{U},1}$ in reconstructing the shadowed areas, especially noticeable for the sample with large $\epsilon=0.10$ in (b).}
     \label{fig:FNO_1_snap}
\end{figure}

This improved wave reconstruction can be illustrated by comparing the reconstructions of the same two exemplary test set samples generated by $\mathcal{M}_{\mathrm{F},1}$ in Figure~\ref{fig:FNO_1_snap} to $\mathcal{M}_{\mathrm{U},1}$ in Figure~\ref{fig:UNet_1_snap}. As depicted in Figure~\ref{subfig:FNO_1_snap_left}, the sample-specific $\mathrm{nL2}_i$ or $\mathrm{SSP}_i$ metrics are only slightly improved, but the ratio  $\tfrac{\mathrm{nL2}_{\mathrm{shad}_i}}{\mathrm{nL2}_{\mathrm{vis}_i}}$ is substantially smaller than observed using $\mathcal{M}_{\mathrm{U},1}$ before. These observations are even more pronounced for the sample with high $\epsilon$ in Figure~\ref{subfig:FNO_1_snap_right}. The $\mathcal{M}_{\mathrm{F},1}$ reduces the error in terms of $\mathrm{nL2}_i$ or $\mathrm{SSP}_i$ by almost half and also produces a more uniform reconstruction between shadowed and visible areas.

\subsubsection{FNO using spatio-temporal radar data}
\label{sec:FNO_multiple_snap}

\begin{figure}[ht!]
\centering
\includegraphics{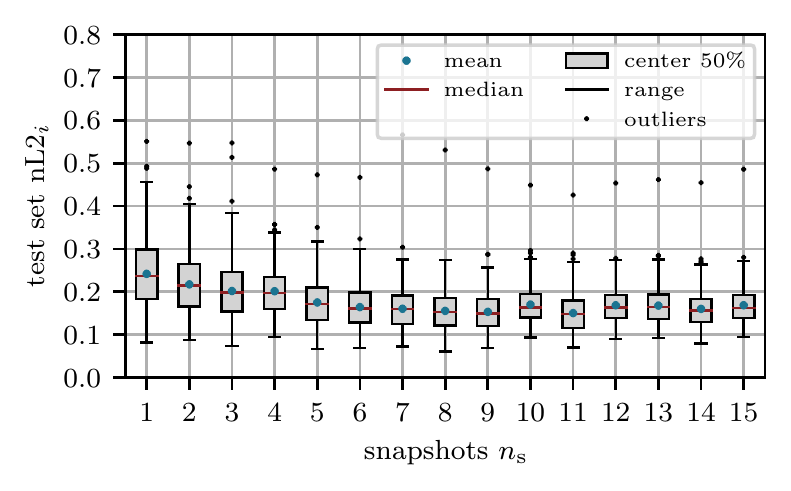}
\vspace{-0.2cm}
\caption{Boxplot depicting the error distribution on the test set, depending on the number of historical radar snapshots $n_\mathrm{s}$ provided to train FNO-based architectures $\mathcal{M}_{\mathrm{F},n_\mathrm{s}}$. The best model performance is achieved for $n_\mathrm{s} = 9$. Afterwards, the errors slightly increase again.}
\label{fig:boxplotFNO}
\end{figure}

Although the FNO-based model $\mathcal{M}_{\mathrm{F},1}$ outperforms the U-Net-based model $\mathcal{M}_{\mathrm{U},1}$, it does not achieve the desired reconstruction quality of $\mathrm{SSP} \le 0.10$ (\ref{cri:2}). To enhance the model performance we analyze the effect of including multiple historical snapshots in each input $\mathbf{x}_i \in \mathbb{R}^{512 \times n_\mathrm{s}}$ for the training of this architecture. Again, 14 additional training runs were conducted, each with an increasing number of $n_\mathrm{s}$. The results, depicted in Figure~\ref{fig:boxplotFNO}, demonstrate an initial improvement in performance for the models $\mathcal{M}_{\mathrm{F},n_\mathrm{s}}$ with increasing $n_\mathrm{s}$ which is slightly less notable than that observed for the U-Net-based models $\mathcal{M}_{\mathrm{U},n_\mathrm{s}}$ in Figure~\ref{fig:boxplotUNet} before. The FNO-based models achieve the best performance for $n_\mathrm{s}=9$ input snapshots, beyond which the mean error slightly increases. 

According to Table \ref{tab:results}, the model $\mathcal{M}_{\mathrm{F},9}$ attains a mean performance of $\mathrm{nL2}=0.153$ on the test set, after 776 training epochs, as depicted by the loss curve in the Figure~\ref{fig:loss_FNO_9_snap}. This error value corresponds to a mean $\mathrm{SSP}=0.076$, fulfilling the \ref{cri:2} of a $\mathrm{SSP \le 0.10}$. However, in comparison to the U-Net-based model $\mathcal{M}_{\mathrm{U},10}$, which achieved a final mean value of $\mathrm{SSP}=0.061$, the performance of $\mathcal{M}_{\mathrm{F},9}$ measured in terms of nL2 or SSP is slightly inferior, even though in the single-snapshot case $\mathcal{M}_{\mathrm{F},1}$ outperformed $\mathcal{M}_{\mathrm{U},1}$. Nevertheless, compared to all investigated models, $\mathcal{M}_{\mathrm{F},9}$ on average achieves the best reconstruction uniformity between shadowed and visible areas indicated by a mean $\tfrac{\mathrm{nL2_\mathrm{shad}}}{\mathrm{nL2}_\mathrm{vis}}=1.381$ on test 

\begin{figure}[ht!]
     \centering
     \begin{subfigure}[b]{0.49\textwidth}
         \centering
         \includegraphics[width=\textwidth]{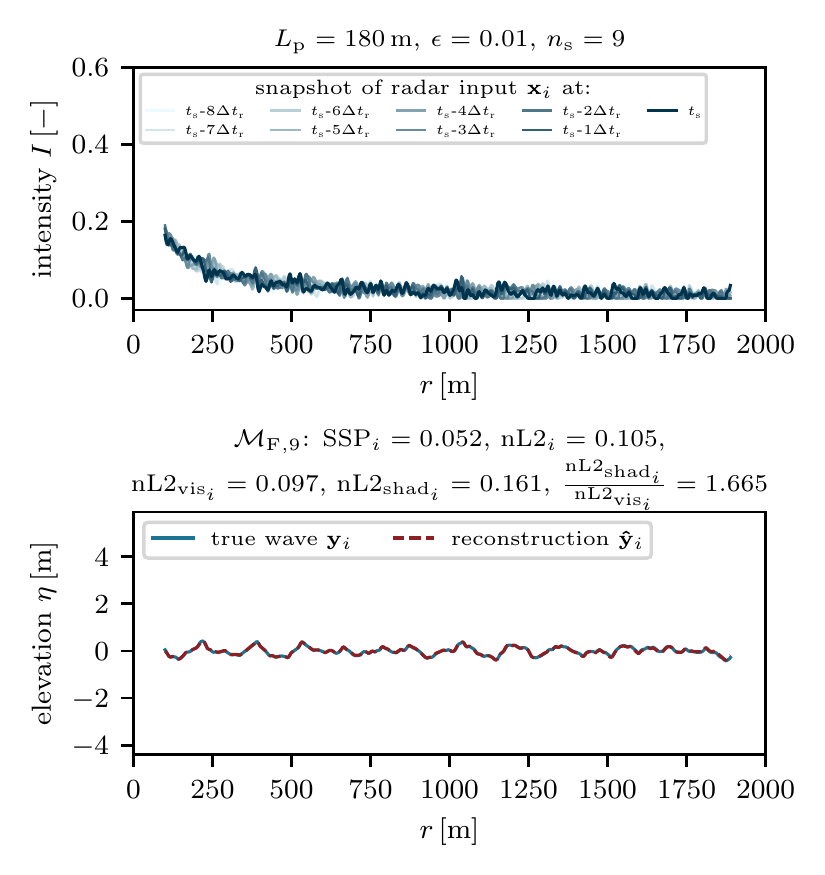}
         \caption{}
         \label{subfig:FNO_9_snap_left}
     \end{subfigure}
     \hfill
     \begin{subfigure}[b]{0.49\textwidth}
         \centering
         \includegraphics[width=\textwidth]{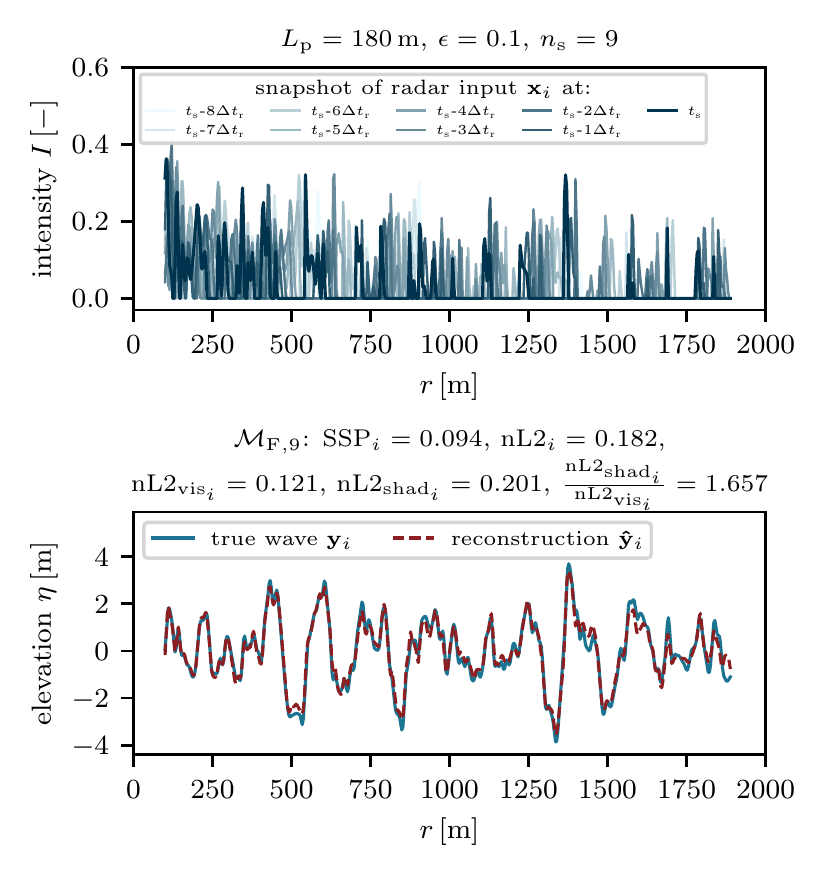}
         \caption{}
         \label{subfig:FNO_9_snap_right}
     \end{subfigure}
     \caption{Two samples from the test set described by the same wavelength $L_\mathrm{p}=180 \, \mathrm{m}$, but different wave steepness $\epsilon$, reconstructed by the FNO-based architecture trained with $n_\mathrm{s}=9$ historical snapshots in the radar input $\mathcal{M}_{\mathrm{F},9}$. Compared to $\mathcal{M}_{\mathrm{F},1}$ a reconstruction improvement is visible for both samples. Moreover, the reconstruction quality on the entire $r$-domain is almost equivalent to the results of $\mathcal{M}_{\mathrm{U},10}$, but especially for the small steepness sample in (a) the error ratio between shadowed and visible areas is remarkably smaller using $\mathcal{M}_{\mathrm{F},9}$ which indicates the potential of a more uniform reconstruction.}
     \label{fig:FNO_9_snap}
\end{figure}

Figure~\ref{fig:FNO_9_snap} shows the reconstructions $\mathbf{\hat{y}_i}$ for the same two exemplary radar inputs $\mathbf{x}_i$ from the test set used before, now generated by the trained FNO-based model $\mathcal{M}_{\mathrm{F},9}$. 
Compared to $\mathcal{M}_{\mathrm{F},1}$ in Figure~\ref{fig:FNO_1_snap} both samples experience an almost similar increase in reconstruction quality measured in terms of the sample-specific $\mathrm{SSP}_i$ and $\mathrm{nL2}_i$ errors.
In addition, these values are comparable to that achieved by $\mathcal{M}_{\mathrm{U},10}$ in Figure~\ref{fig:UNet_10_snap}. However, for the sample with small $\epsilon=0.01$ in Figure~\ref{subfig:FNO_9_snap_left}, $\mathcal{M}_{\mathrm{F},9}$ generates a more balanced reconstruction than $\mathcal{M}_{\mathrm{U},10}$, as reflected by the reduction of $\tfrac{\mathrm{nL2_{\mathrm{shad}_i}}}{\mathrm{nL2}_{\mathrm{vis}_i}}$ from 2.201 to 1.665 for this individual sample. For the higher-steepness sample in Figure~\ref{subfig:FNO_9_snap_right}, the increase of reconstruction uniformity given by $\tfrac{\mathrm{nL2_{\mathrm{shad}_i}}}{\mathrm{nL2}_{\mathrm{vis}_i}}$ is less significant but still present.

\subsection{Comparative discussion}

The aforementioned visual observations described for Figures \ref{fig:UNet_1_snap}, \ref{fig:UNet_10_snap}, \ref{fig:FNO_1_snap} and \ref{fig:FNO_9_snap} have been limited to the examination of only two exemplary samples from the test set, both described by peak wavelength $L_\mathrm{p}=180 \, \mathrm{m}$ and either steepness $\epsilon=0.01$ or $\epsilon=0.10$. To avoid any possible incidental observations, the generalization of the error values needs to be examined. This can be achieved by plotting sample-specific error values such as $\mathrm{nL2}_i$ against each sample's describing combination of peak wavelength $L_\mathrm{p}$ and steepness $\epsilon$ for all $N_\mathrm{test}=624$ test set samples reconstructed using the U-Net-based models $\mathcal{M}_{\mathrm{U},1}$ and $\mathcal{M}_{\mathrm{U},10}$ or the FNO-based models $\mathcal{M}_{\mathrm{F},1}$ and $\mathcal{M}_{\mathrm{F},9}$. 

\subsubsection{Discussion of overall reconstruction quality}

\begin{figure}[ht!]
     \centering
     \begin{subfigure}[b]{0.48\textwidth}
         \centering
         \includegraphics[width=\textwidth]{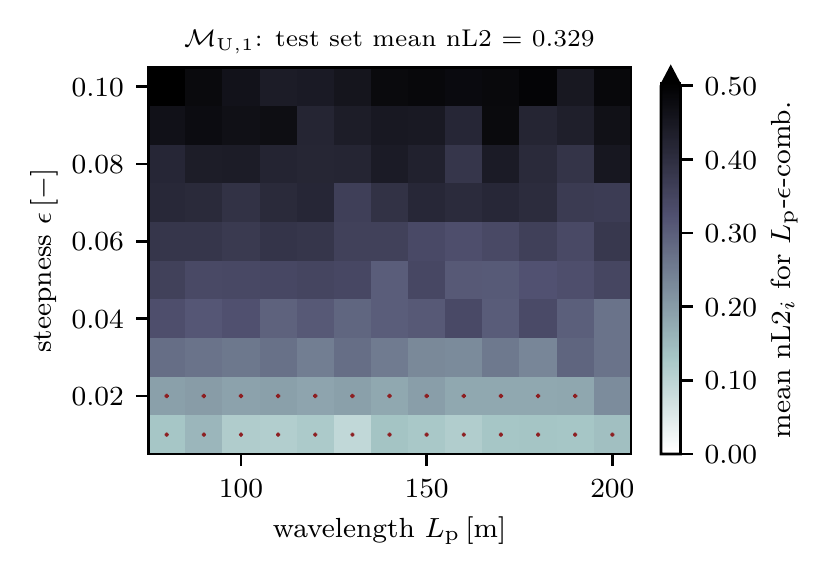}
         
         \caption{}
         \label{subfig:UNet1}
     \end{subfigure}
     \hfill
    \begin{subfigure}[b]{0.48\textwidth}
         \centering
         \includegraphics[width=\textwidth]{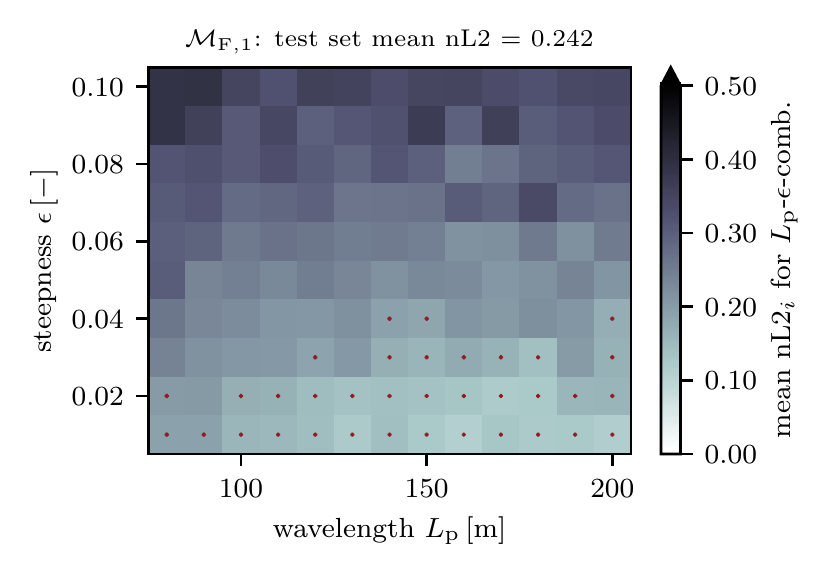}
         \caption{}
         \label{subfig:FNO1}
     \end{subfigure}
     \begin{subfigure}[b]{0.48\textwidth}
         \centering
         \includegraphics[width=\textwidth]{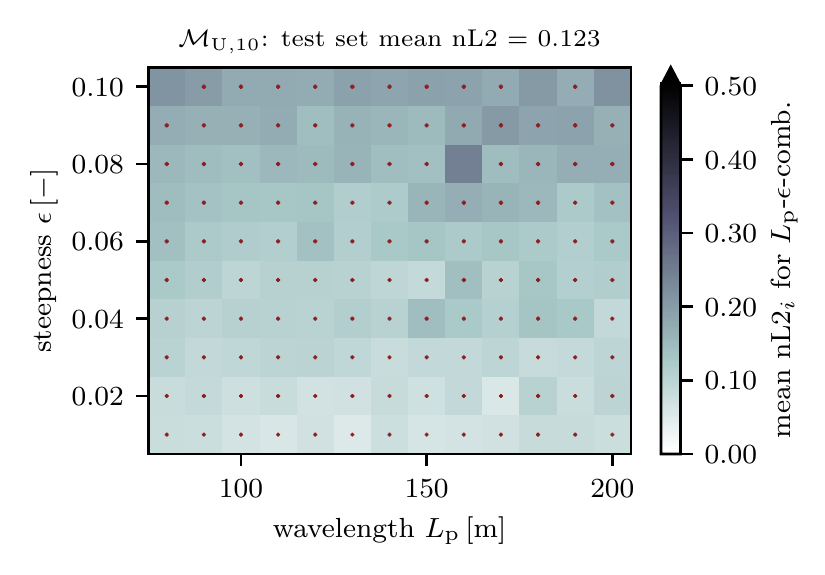}
         \caption{}
         \label{subfig:UNet10}
     \end{subfigure}
     \hfill
     \begin{subfigure}[b]{0.48\textwidth}
         \centering
         \includegraphics[width=\textwidth]{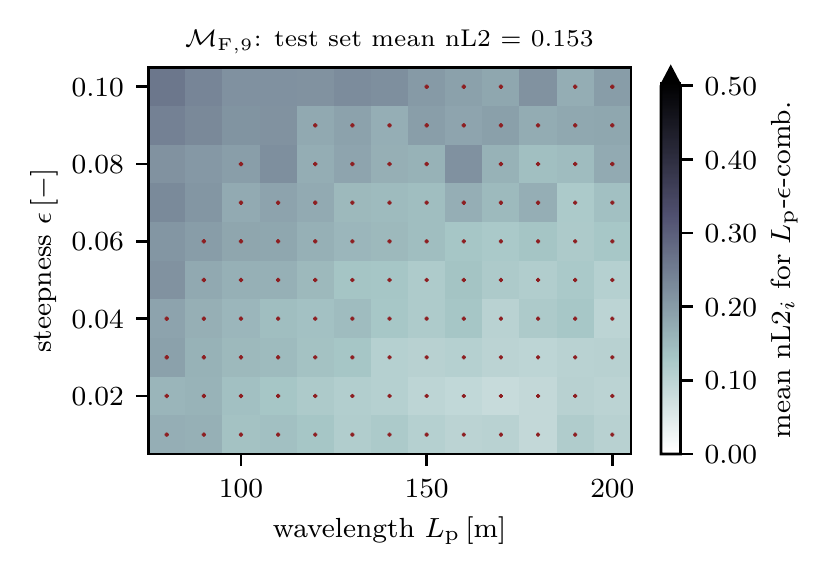}
         \caption{}
         \label{subfig:FNO9}
     \end{subfigure}
     \caption{Error surfaces generalizing the previous observations for sample-specific errors $\mathrm{nL2}_i$ of the four investigated models $\mathcal{M}$ depending on the $L_\mathrm{p}$-$\epsilon$-combination of the samples from the test set. Red dots indicate parameter combinations that meet the \ref{cri:2} of reconstruction errors $\mathrm{SSP}_i \le 0.10$. The upper subplots illustrate the result of (a) the U-Net-based model and (b) the FNO-based model, both trained with only one radar snapshot ($n_\mathrm{s}=1$) in each input $\mathbf{x}_i \in \mathbb{R}^{n_r \times 1}$. The same architectures were trained with multiple historic radar snapshots in each input  $\mathbf{x}_i \in \mathbb{R}^{n_r \times n_\mathrm{s}}$, as demonstrated in the lower subplots, where (c) shows the U-Net-based model trained with $n_\mathrm{s}=10$ and (d) the FNO-based model trained with $n_\mathrm{s}= 9$.}
     \label{fig:generalization}
\end{figure}

Figure~\ref{fig:generalization} illustrates the reconstruction error as the mean $\mathrm{nL2}_i$ value across 4-5 samples available for each specific $L_\mathrm{p}$-$\epsilon$-combination included in the test set. Additionally, red dots in the cell centers indicate the combinations that achieved a mean $\mathrm{SSP}_i \le 0.10$ (\ref{cri:2}).

\begin{figure}[ht!]
\centering
\includegraphics{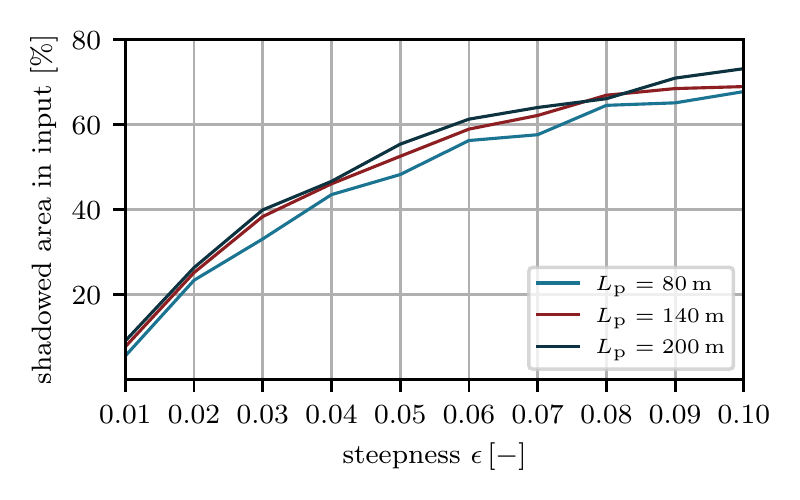}
\vspace{-0.2cm}
\caption{Graphs visualizing the average proportion of each input $\mathbf{x}_i$ being affected by shadowing modulation in dependency of the samples wave steepness values $\epsilon=0.01-0.10$ for the shortest, one medium and the longest peak wavelength $L_\mathrm{p}$ occurring in the test set.}
\label{fig:shadowing_over_steep}
\end{figure}

Subfigure~\ref{subfig:UNet1} confirms the findings presented in Section \ref{sec:UNet_single_snap} for the U-Net-based model $\mathcal{M}_{\mathrm{U},1}$ trained with one radar snapshot ($n_\mathrm{s}=1$) in each input $\mathbf{x}_i$. The  errors between the true $\mathbf{y}_i$ and reconstructed wave output $\mathbf{\hat{y}}_i$ increase with increasing steepness $\epsilon$ and thus with increasing wave height. Moreover, we now observe that this effect occurs almost independent of the peak wavelength $L_\mathrm{p}$ of each sample. For samples described by $\epsilon > 0.02$, $\mathcal{M}_{\mathrm{U},1}$ fails to meet the \ref{cri:2} as the corresponding errors exceed $\mathrm{SSP}_i$ values of $0.10$. This is attributable to the geometrical radar imaging problem demonstrated in Figure~\ref{fig:tilt+shadowing}, showing that the increase in wave height caused by increased $\epsilon$ results in more and larger shadowed areas. Figure~\ref{fig:shadowing_over_steep} demonstrates that the occurrence of shadowing mainly increases with increasing $\epsilon$ and is less influenced by $L_\mathrm{p}$. While $\epsilon=0.01$ on average only causes around $10 \%$, $\epsilon=0.10$ instead causes approximately $70-75 \%$ of each input $\mathbf{x}_i$ being affected by shadowing modulation. This results in areas along the spatial range $r$ containing zero-valued intensities that complicate the radar inversion task.

Understanding the challenges faced by model $\mathcal{M}_{\mathrm{U},1}$ in reconstructing shadowed areas, requires revisiting the U-Net's local mode of operation, outlined in Section \ref{sec:U-Net methodology}, and the exemplary radar input depicted in the upper panel of Figure~\ref{subfig:UNet_1_snap_right}. Due to shadowing, numerous local areas exhibit zero-intensities covering up to approximately $200 \, \mathrm{m}$, especially for greater distances from the radar system. However, the kernels in the first convolutional layer with a kernel size of $s_\mathrm{k}=5$ only cover a domain of $s_\mathrm{k} \cdot \Delta r = 17.5 \, \mathrm{m}$ while being shifted across the input feature map in a step-wise manner. While the U-Net's translational equivariance property is useful for translating radar intensities to wave surface elevation regardless of their spatial location, it thus also causes kernels to be shifted across large areas with zero input only, which cannot be processed in a meaningful way. 
Although the pooling layers subsequently reduce the dimension of feature maps, resulting in an increased ratio of kernel size to feature size, the problem of radar inversion can be assumed to be based on the mapping of individual pixel values, known as low-level features. These features are learned in the early layers of a CNN-based network \citep{Zeiler2013}. Accordingly, the initial stages of the U-Net-based architecture are more important for our task than for its original purpose of image segmentation \citep{Ronneberger2015} that is based on mid- to high-level features extracted in the later layers. For this reason, we face problems applying $\mathcal{M}_{\mathrm{U},1}$ for reconstruction, as important kernels in the early layers receive a significant amount of sparse, not valuable content. Although increasing the kernel size $s_\mathrm{k}$ is a theoretically possible solution, doing so would compromise the U-Net's local key property. Moreover, when processing two-dimensional surfaces with 2D convolutional kernels in future research, it would result in a quadratic increase in the number of weights, leading to computational issues.

For this reason, the approach of providing $n_\mathrm{s}=10$ consecutive radar snapshots governed according to \ref{cri:3} for the training of U-Net-based model $\mathcal{M}_{\mathrm{U},10}$ in Section \ref{sec:UNet_multiple_snap} more effectively accounts for the sparsity in the input data. The upper panel of Figure~\ref{subfig:UNet_10_snaps_right} demonstrated the presence of input information across the majority of the $r$-domain. The wave surfaces undergo shape variations while travelling towards the radar due to differing phase velocities of their components caused by dispersion. This results in a different part of the radar surface being shadowed or visible at each time step and seem to allow to capture more information about the wave on average, as the reconstruction quality significantly improves compared to $\mathcal{M}_{\mathrm{U},1}$. Therefore we infer that the spatial and temporal shifts of the additional radar intensities acquired at $t_j = \sum_{j=0}^{n_\mathrm{s}-1} t_\mathrm{s}- j \Delta t_\mathrm{r}$ can be compensated successfully. This may be attributed to the fact that each kernel applied to the input has its own channel for each snapshot, allowing for separate processing to counterbalance the shift first, followed by the addition of results to one feature map utilized as part of the input for the next layer.
The improved reconstruction observed for $\mathcal{M}_{\mathrm{U},10}$ is further supported by its performance generalization shown in Figure~\ref{subfig:UNet10}.  Compared to Figure~\ref{subfig:UNet1}, the mean nL2 error is substantially smaller and sample-specific reconstruction errors $\mathrm{nL2}_i $ are more evenly distributed across the $L_\mathrm{p}$-$\epsilon$-space, resulting in a satisfactory $\mathrm{SSP}_i $ value (\ref{cri:2}) for almost all samples. Although there is still a slight increase in the error for samples with higher $L_\mathrm{p}$ and $\epsilon$, the proposed model $\mathcal{M}_{\mathrm{U},10}$ can accurately reconstruct samples with varying wave characteristics and degrees of shadowing, thus supporting \ref{hyp:1} and \ref{hyp:3}.

Motivated by the inherent patterns in wave data and the successful application of the Fourier neural operator (FNO) to systems exhibiting certain periodic properties, we conducted a comparative analysis of the global inductive bias of this network architecture with the local inductive bias of the CNN-based U-Net. As discussed in Section \ref{sec:FNO_single_snap}, our observations indicate that the FNO-based model $\mathcal{M}_{\mathrm{F},1}$ trained with only one snapshot ($n_\mathrm{s}=1$) outperforms the U-Net-based $\mathcal{M}_{\mathrm{U},1}$ in reconstructing shadowed areas in the input, as evidenced for example by comparing the reconstruction in Figure~\ref{subfig:FNO_1_snap_right} to \ref{subfig:UNet_1_snap_right}. This observation is generalizable to the entire test data set, as shown in Figure~\ref{subfig:FNO1}. Although errors in the FNO error surface still increase with higher steepness $\epsilon$ and consequently with an increase in the percentage of shadowing according to Figure~\ref{fig:shadowing_over_steep}, the increase is much less severe than that obtained by $\mathcal{M}_{\mathrm{U},1}$ shown in Figure~\ref{subfig:UNet1}.

The improved ability of the FNO-based model $\mathcal{M}_{\mathrm{F},1}$ in reconstructing shadowed areas from a single-snapshot input can be attributed to its mode of operation outlined in Section \ref{sec:FNO methodology}. Although the latent representation $v_0$ in Figure~\ref{fig:FNO} is usually not explicitly known, we can infer that the layer $P$ with $n_\mathrm{s}=1$ input nodes and $n_\mathrm{w}$ output nodes only performs linear transformations to each radar input $\mathbf{x}_i$. As the radar inputs exhibit kinks at the transitions from visible to shadowed areas, $v_0$ will have similar characteristics along the range direction. These transitions result in peaks for specific wavenumbers $k$ in the spectrum $F(k)$. However, the desired wave outputs $\mathbf{y}_i$ of the training data samples possess smooth periodic properties, without peaks at the kink-related wavenumbers in $F_\mathbf{y}(k)$. Since the $R_j$ matrices in the Fourier layers scale the radar input spectrum to the wave output spectrum, they learn small coefficients for the corresponding entries to reduce the peaks. Therefore, the FNO's global inductive bias, combined with the data structure of wave surfaces, can efficiently correct sparse, shadowed regions in spectral space, resolving the issue of insufficient local information for reconstruction that arises with the U-Net-based model $\mathcal{M}_{\mathrm{U},1}$ in Euclidean space. Thus it can be also stated that the FNO explicitly hard-encodes prior knowledge about physical wave properties through its network structure and thus can be assumed to be a \textit{physics-guided design of architecture} \citep[cf.][]{Willard2022, Wang2023} for our problem. 

Despite the better performance of the FNO-based model $\mathcal{M}_{\mathrm{F},1}$ compared to $\mathcal{M}_{\mathrm{U},1}$ in reconstructing shadowed radar inputs that already supports \ref{hyp:2}, the red dots in Figure~\ref{subfig:FNO1} still reveal that most of the test set samples fail to meet the \ref{cri:2} of $\mathrm{SSP}_i \le 0.10$. However, this issue was resolved by training a FNO-based model $\mathcal{M}_{\mathrm{F},9}$ with $n_\mathrm{s}=9$ historical radar snapshots in each input $\mathbf{x}_i$. This was demonstrated for the two test set examples in Figure~\ref{fig:FNO_9_snap} and is generalized in Figure~\ref{subfig:FNO9}. We observe from that Figure, that the slightly higher mean error across the entire test set of $\mathcal{M}_{\mathrm{F},9}$ compared to $\mathcal{M}_{\mathrm{U},10}$, is primarily caused by the individual errors $\mathrm{nL2}_i$ of samples with low steepness $\epsilon$ or short wavelengths $L_\mathrm{p}$. It is worth noting, that the observed minimal increase in errors for short wavelengths cannot be attributed to a truncation at an insufficient number of Fourier series modes $n_\mathrm{m}$ in the Fourier layers. In this work, $n_\mathrm{m}$ is determined as 64 and the spectral representation is discretized by $\Delta k = \tfrac{2 \pi}{n_r \cdot \Delta r} = 0.00351 \, \mathrm{m}^{-1}$. The highest peak wavenumber of $k_\mathrm{p}= 0.0785 \, \mathrm{m}^{-1}$ in our data set is reached for samples with $L_\mathrm{p}= 80 \, \mathrm{m}$. The spectral density around $k_\mathrm{p}$ has decayed almost completely at $k_\mathrm{filt} = n_\mathrm{m} \cdot \Delta k = 0.2264\, \mathrm{m}^{-1}$, such that no important wave components are filtered out, as is visualized in the Figure~\ref{fig:JONSWAP_Lps}. Therefore, the small unequal tendency in error distribution achieved by $\mathcal{M}_{\mathrm{F},9}$ in Figure~\ref{subfig:FNO9}  for samples described by different $L_\mathrm{p}$-$\epsilon$, is likely caused by other factors than by an unsuitable network hyperparameter $n_\mathrm{m}$. Moreover, we observed in the loss curve shown in Figure~\ref{fig:loss_FNO_9_snap} that further training for more than 800 epochs could potentially improve the model's performance, whereas the best performance on the test set for $\mathcal{M}_{\mathrm{U},10}$ seems to be already reached, as the model begins to overfit the training data, as depicted in Figure~\ref{fig:loss_UNet_1_snap}. \\

\subsubsection{Discussion of reconstruction uniformity}

\begin{figure}[ht!]
     \centering
     \begin{subfigure}[b]{0.48\textwidth}
         \centering
         \includegraphics[width=\textwidth]{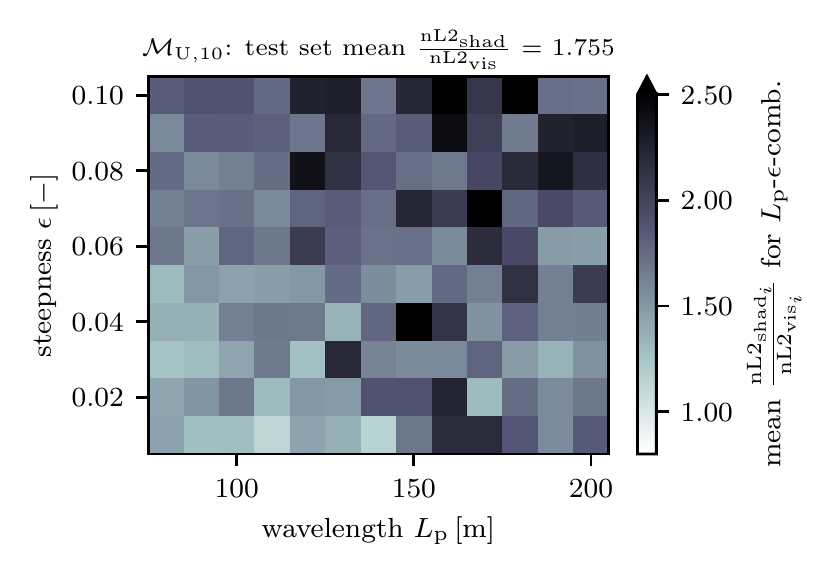}
         \caption{}
         \label{subfig:UNet10_ratio}
     \end{subfigure}
     \hfill
    \begin{subfigure}[b]{0.48\textwidth}
         \centering
         \includegraphics[width=\textwidth]{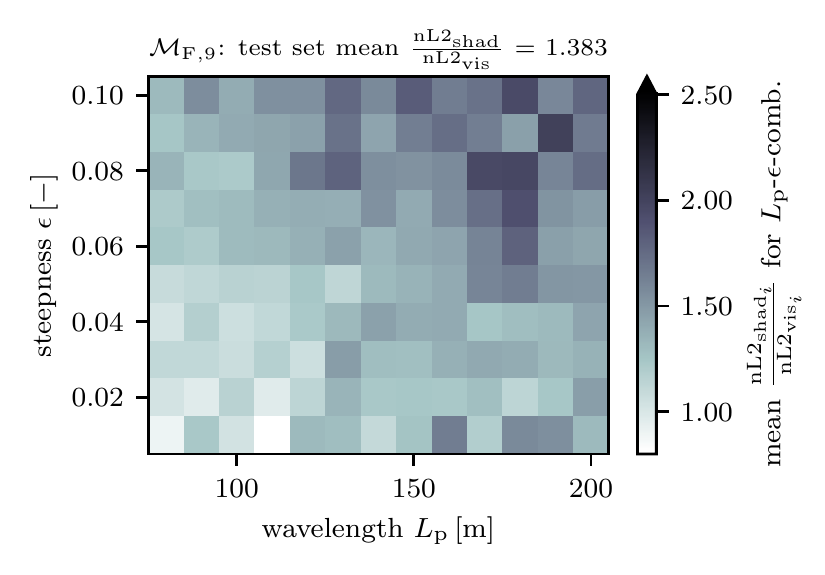}
         \caption{}
         \label{subfig:FNO9_ratio}
     \end{subfigure}
     \caption{Error surfaces depicting the ratio $\tfrac{\mathrm{nL2}_{\mathrm{shad}_i}}{\mathrm{nL2}_{\mathrm{vis}_i}}$ between the reconstruction quality achieved on shadowed and visible areas depending on the specific $L_\mathrm{p}$-$\epsilon$-combination of the samples from the test set. The individual cell entries display the mean ratio across the 4-5 samples available for each specific parameter combination. The uniformity of the reconstructions achieved by the U-Net-based model $\mathcal{M}_{\mathrm{U},10}$ in (a) thus is compared to the one achieved by the FNO-based model $\mathcal{M}_{\mathrm{F},9}$ in (b). }
     \label{fig:error_surfaces_ratio}
\end{figure}

So far the generalization of the reconstruction quality has been evaluated based on samples-specific $\mathrm{nL2}_i$ or $\mathrm{SSP}_i$ values across the entire spatial $r$-domain. However, Table \ref{tab:results} indicates that the FNO-based model $\mathcal{M}_{\mathrm{F},9}$ achieves a more uniform reconstruction between shadowed and visible areas. This is demonstrated by the mean ratio of $\tfrac{\mathrm{nL2}_\mathrm{shad}}{\mathrm{nL2}_\mathrm{vis}}=1.381$ across all samples in the test set, while the U-Net-based model $\mathcal{M}_{\mathrm{U},10}$ still struggles with reconstructing shadowed areas as inferred by its $\tfrac{\mathrm{nL2}_\mathrm{shad}}{\mathrm{nL2}_\mathrm{vis}}=1.755$. Therefore, the $\tfrac{\mathrm{nL2}_{\mathrm{shad}_i}}{\mathrm{nL2}_{\mathrm{vis}_i}}$-ratio error distribution is displayed in Figure~\ref{fig:error_surfaces_ratio} for each test set sample based on their $L_\mathrm{p}$-$\epsilon$-combination. The model $\mathcal{M}_{\mathrm{U},10}$ generates an error surface shown in Figure~\ref{subfig:UNet10_ratio} that exhibits broadly varying levels of uniformity in the reconstruction even for samples with neighbouring $L_\mathrm{p}$-$\epsilon$-combinations. In some cases, the reconstruction errors in shadowed areas exceed those in visible areas by more than 2.5 times.  This undesired effect is much less pronounced for the FNO-based model $\mathcal{M}_{\mathrm{F},9}$, as a comparison with Figure~\ref{subfig:FNO9_ratio} reveals.

\subsubsection{Final comparison}

For a final evaluation, either the general reconstruction quality nL2 can be chosen  as the main performance criterion, which in our case would argue for the selection of the U-Net-based model $\mathcal{M}_{\mathrm{U},10}$, or instead the uniformity of the reconstruction indicated by $\tfrac{\mathrm{nL2}_\mathrm{shad}}{\mathrm{nL2}_\mathrm{vis}}$, which would argue for the FNO-based model $\mathcal{M}_{\mathrm{F},9}$. This decision should be made based on the application case. If the ML-reconstructed wave surface is intended to be used as an initial condition for subsequent prediction with the HOS method, we would expect a more uniform reconstruction to represent a more physical result, and consequently, the FNO-based reconstruction to be less likely to affect the subsequent wave prediction in a negative way. Moreover, we observed that the global approach of the FNO-based models would allow for a reasonably more meaningful reconstruction of shadowed areas even with fewer historical radar snapshots $n_\mathrm{s}$ contained in each input $\mathbf{x}_i$. This is not necessarily the case using the U-Net-based models. 

Besides, the FNO-based model $\mathcal{M}_{\mathrm{F},9}$ in this work allows for much faster inference speed than the U-Net-based $\mathcal{M}_{\mathrm{U},10}$, even though $\mathcal{M}_{\mathrm{F},9}$ is constructed as a custom implementation and contains more weights compared to $\mathcal{M}_{\mathrm{U},10}$, which uses standard layers from the PyTorch library that are in addition probably optimized. More specifically, using the hardware specifications outlined in Section \ref{sec:Training+evaluation}, our $\mathcal{M}_{\mathrm{F},9}$ is able to generate reconstructions $\mathbf{\hat{y}}_i$ for an input sample $\mathbf{x}_i$ in an average time of $1.9 \cdot 10^{-3} \, \mathrm{s}$ which is approximately 20 times faster than the average time of $3.7  \cdot 10^{-2}  \, \mathrm{s}$ required by $\mathcal{M}_{\mathrm{U},10}$ for the same task.

%%% Conclusions %%%
\section{Conclusion}
\label{sec:conclusion}

This work introduces a novel machine learning-based approach for the phase-resolved reconstruction of ocean wave surface elevations from sparse radar measurements. To evaluate the performance of our approach, we generate synthetic nonlinear wave surface data for a wide range of sea states and corresponding radar surface data by incorporating both tilt- and shadowing modulation mechanisms. Two neural network architectures based on the U-Net or the Fourier neural operator are trained, both provided with varying amounts of spatio-temporal radar surface measurement input.

Our results and discussion indicate that both models are capable of producing high-quality wave surface reconstructions with average errors below $\mathrm{SSP}\le 0.10$ when trained with a sufficient amount of $n_\mathrm{s}=10$ or $9$ consecutive radar snapshots. Furthermore, both models generalize well across different sea states. On average, the U-Net-based model achieves slightly smaller errors across the entire spatial domain of each reconstructed wave sample, while the FNO-based model produces a more uniform wave reconstruction between areas that were shadowed and visible in the corresponding radar input.
This observation is further confirmed by the edge case of instantaneous inversion, i.e. if the networks are trained with only a single radar snapshot in each input. The weakness in the reconstruction of shadowing-affected areas of the U-Net-based model can be attributed to the local operation of the network architecture, where its small convolutional kernels do not receive processable information when shifted across shadowed input areas with zero intensities only. The problem can be circumvented using the FNO-based network that learns a global mapping between radar input and wave output in the Fourier space. This network structure already encodes prior physical knowledge about the periodic data structure apparent in ocean waves and is therefore possibly better suited for our use case.

Our findings suggest that the FNO-based network may offer additional advantages, especially concerning smaller training datasets and noisy input radar data. Furthermore, future research could delve into the reconstruction of two-dimensional ocean wave surfaces, as the FNO network can also be implemented using 2D-FFTs. However,  due to the different propagation directions of the component waves in short-crested, two-dimensional sea states, we anticipate a potential degradation in the reconstruction performance compared to the one-dimensional scenario explored in this study. This performance degradation could be mitigated through appropriate countermeasures, such as employing FNOs with increased capacity, conducting longer training runs, or applying suitable regularization techniques.

Moreover, the current methodology solely relies on synthetic radar input and the corresponding wave output data. Although it can be presumed that the HOS method generates wave surfaces that exhibit a reasonable degree of realism, radar imaging mechanisms for marine X-Band radar are not yet fully understood, such that state-of-the-art radar models are associated with higher uncertainties. Consequently, on the one hand, a model trained solely on synthetic radar-wave data pairs cannot be applied for inference using real-world radar data. On the other hand, the acquisition of real-world radar-wave pair samples to train the neural networks is associated with high operational costs due to the necessity of deploying a dense grid of buoys for capturing wave snapshots. These data issues currently limit the application of the developed machine-learning-based reconstruction approach for real-world applications. In future research, we endeavour to tackle this issue through two opportunities: Firstly, we aim to enhance the realism of synthetic radar data models to improve their accuracy. Alternatively, we intend to investigate the feasibility of physics-informed learning approaches as a tool to overcome the challenges associated with measuring real-world high-resolution wave output data.

\clearpage

%%% Appendix %%%
\appendix
\section{Influence of neural network hyperparameters}
\label{sec:Appendix-Hyperparameters}

To mitigate the high cost of obtaining a larger data set, a four-fold cross-validation approach with an independent test set was utilized for finding the network hyperparameters, as recommended for example by \cite{Raschka2018}. The data set of $N=3120$ samples was divided into a fixed and independent test set comprising $20\%$ or $N_\mathrm{test}=624$ samples, with the remaining 2496 samples partitioned into four equal-sized parts based on the governing sea state parameters $(L_\mathrm{p}, \epsilon)$ using a stratified data split technique to ensure equal representation of each wave characteristic in the resulting subsets. During each cross-validation step, one part with $N_\mathrm{val}=624$ samples was used as the validation set, and the remaining three parts with $N_\mathrm{train}=1872$ samples constituted the training set.

Tables \ref{tab:hyerparametersUNet} and \ref{tab:hyperparametersFNO} present the results of the four-fold cross-validation hyperparameter studies for the U-Net- and FNO-based architectures. For both network types, the same fixed test set was excluded from this investigation. The metrics ($\mathrm{nL2}, \, \frac{\mathrm{nL2}_\mathrm{shad}}{\mathrm{nL2}_\mathrm{vis}}, \, \mathrm{SSP}$) and the number of epochs necessary to attain the best performance represent average values across all four folds.

\begin{table}[!htp]
\centering
\caption{Results of the hyperparameter study for the U-Net-based architecture. Each investigated architecture is characterized by a depth $n_\mathrm{d}$, a kernel size $s_\mathrm{k}$ and an approach for the number of convolutional kernels $ n_\mathrm{k}$ in each layer. The approaches for $n_\mathrm{k}$ explored doubling the number of kernels per layer with increasing encoder depth (and in reverse halving in the decoder) and another keeping $n_\mathrm{k}$  the same in all convolutional layers. The performance, measured by the mean nL2 value on validation set displays only slight variations among architectures, indicating that the U-Net is not highly sensitive to these changes. Nevertheless, we select the best model according to the validation nL2, that highlighted in blue, as our $\mathcal{M}_{\mathrm{U},1}$.}
\begin{scriptsize}
\begin{tabular}{llcrrcccccc}
\toprule
\toprule
\multicolumn{3}{c}{U-Net hyperparameters} & \multicolumn{2}{c}{} & \multicolumn{2}{c}{$\mathrm{nL2}$} & \multicolumn{2}{c}{$\frac{\mathrm{nL2}_\mathrm{shad}}{\mathrm{nL2}_\mathrm{vis}}$}& \multicolumn{2}{c}{SSP} \\
\cmidrule(rl){1-3} \cmidrule(rl){6-7} \cmidrule(rl){8-9} \cmidrule(rl){10-11}
depth $n_\mathrm{d}$ & kernels each layer $n_\mathrm{k}$ & k. size $s_\mathrm{k}$ & \#weights & epochs & train & val & train & val & train & val  \\ 
\midrule
\midrule
\multirow{2}{*}{3} & 
\multirow{2}{*}{$[8, 16, 32]$} & 
      3 &  12,401 & 785 & 0.392 & 0.395 & 2.124 & 2.119 & 0.208 & 0.210 \\
    && 5 &  18,305 & 791 & 0.360 & 0.367 & 2.512 & 2.526 & 0.189 & 0.193\vspace{0.12cm} \\

\multirow{2}{*}{4} & 
\multirow{2}{*}{$[8, 16, 32, 64]$} & 
      3 &  48,433 & 645 & 0.352 & 0.371 & 2.475 & 2.527 & 0.185 & 0.195 \\
    && 5 &  72,769 & 189 & 0.347 & 0.361 & 2.536 & 2.575 & 0.182 & 0.189\vspace{0.12cm} \\

\multirow{2}{*}{5} & 
\multirow{2}{*}{$[8, 16, 32, 64, 128]$} & 
      3 & 192,177 & 278 & 0.342 & 0.360 & 2.601 & 2.644 & 0.179 & 0.189 \\
    && 5 & 290,241 & 154 & 0.303 & 0.344 & 2.276 & 2.438 & 0.158 & 0.179\vspace{0.12cm} \\

\multirow{2}{*}{6} & 
\multirow{2}{*}{$[8, 16, 32, 64, 128, 256]$} & 
      3 & 766,385 & 223 & 0.336 & 0.356 & 2.557 & 2.601 & 0.176 & 0.186 \\
    && 5 &1,159,361 &  78 & 0.326 & 0.356 & 2.365 & 2.452 & 0.170 & 0.186\vspace{0.05cm} \\
\midrule

\multirow{2}{*}{3} & 
\multirow{2}{*}{$[16, 32, 64]$} & 
      3 &  49,121 & 779 & 0.388 & 0.392 & 2.149 & 2.144 & 0.206 & 0.208 \\
    && 5 &  72,705 & 675& 0.353 & 0.366 & 2.543 & 2.582 & 0.185 & 0.193\vspace{0.12cm} \\

\multirow{2}{*}{4} & 
\multirow{2}{*}{$[16, 32, 64, 128]$} & 
      3 & 192,865 & 310 & 0.356 & 0.371 & 2.439 & 2.479 & 0.187 & 0.195 \\
    && 5 & 290,177 & 99  & 0.349 & 0.364 & 2.520 & 2.551 & 0.183 & 0.191\vspace{0.12cm} \\

\multirow{2}{*}{5} & 
\multirow{2}{*}{$[16, 32, 64, 128, 256]$} & 
      3 & 767,073 & 140 & 0.343 & 0.365 & 2.490 & 2.532 & 0.181 & 0.192 \\
    && 5 & 1,159,297 & 76 & 0.298 & 0.342 & 2.364 & 2.476 & 0.155 & 0.178\vspace{0.12cm}  \\

\multirow{2}{*}{6} & 
\multirow{2}{*}{$[16, 32, 64, 128, 256, 512]$} & 
      3 & 3,062,369 & 225 & 0.338 & 0.358 & 2.568 & 2.611 & 0.177 & 0.188 \\ 
    && 5 & 4,634,241 & 124 & 0.346 & 0.366 & 2.419 & 2.487 & 0.183 & 0.194\vspace{0.05cm}  \\
\midrule

\multirow{2}{*}{3} & 
\multirow{2}{*}{$[32, 32, 32]$} & 
       3 &  38465 & 778 & 0.388 & 0.392 & 2.135 & 2.137 & 0.206 & 0.208 \\
    && 5 &  52865 & 724 & 0.351 & 0.363 & 2.581 & 2.611 & 0.185 & 0.191\vspace{0.12cm} \\

\multirow{2}{*}{4} & 
\multirow{2}{*}{$[32, 32, 32, 32]$} &
      3 &  49,825 & 569 & 0.350 & 0.364 & 2.555 & 2.583 & 0.184 & 0.192 \\
    && 5 &  70,369 & 199 & 0.345 & 0.358 & 2.631 & 2.663 & 0.181 & 0.188\vspace{0.12cm} \\

\multirow{2}{*}{5} & 
\multirow{2}{*}{$[32, 32, 32, 32, 32]$} &
      3 &  61,185 & 224 & 0.344 & 0.357 & 2.654 & 2.663 & 0.180 & 0.187 \\
    && 5 & \cellcolor{DynCyan!40}87,873 & \cellcolor{DynCyan!40}183 & \cellcolor{DynCyan!40}0.310 & \cellcolor{DynCyan!40}0.341 & \cellcolor{DynCyan!40}2.536 & \cellcolor{DynCyan!40} 2.659 & \cellcolor{DynCyan!40}0.161 & \cellcolor{DynCyan!40}0.177\vspace{0.12cm} \\

\multirow{2}{*}{6} & 
\multirow{2}{*}{$[32, 32, 32, 32, 32, 32]$} &
      3 & 72,545  & 395 & 0.350 & 0.362 & 2.548 & 2.562 & 0.184 & 0.191 \\
    && 5 & 105,377 & 217 & 0.333 & 0.350 & 2.693 & 2.741 & 0.174 & 0.183\vspace{0.05cm} \\
    \midrule

\multirow{2}{*}{3} & 
\multirow{2}{*}{$[64, 64, 64]$} & 
      3 & 152,705 & 739 & 0.383 & 0.389 & 2.186 & 2.189 & 0.203 & 0.206 \\
    && 5 & 210,177 & 394 & 0.351 & 0.365 & 2.539 & 2.568 & 0.185 & 0.192\vspace{0.12cm} \\ 

\multirow{2}{*}{4} & 
\multirow{2}{*}{$[64, 64, 64, 64]$} & 
     3 & 197,953 & 333 & 0.351 & 0.367 & 2.482 & 2.524 & 0.184 & 0.193 \\
   && 5 & 280,001 & 120 & 0.343 & 0.359 & 2.611 & 2.646 & 0.179 & 0.187\vspace{0.12cm} \\

\multirow{2}{*}{5} & 
\multirow{2}{*}{$[64, 64, 64, 64, 64]$} & 
     3 & 243,201 & 191 & 0.342 & 0.356 & 2.687 & 2.694 & 0.179 & 0.186 \\
   && 5 & 349,825 & 182 & 0.316 & 0.348 & 2.482 & 2.603 & 0.165 & 0.182\vspace{0.12cm} \\

\multirow{2}{*}{6} & 
\multirow{2}{*}{$[64, 64, 64, 64, 64, 64]$} & 
     3 & 288,449 & 255 & 0.345 & 0.362 & 2.532 & 2.568 & 0.181 & 0.190 \\
   && 5 & 419,649 & 138 & 0.323 & 0.349 & 2.598 & 2.684 & 0.168 & 0.182\vspace{0.05cm} \\
\bottomrule
\bottomrule
\end{tabular}
\end{scriptsize}
\label{tab:hyerparametersUNet}
\end{table}

\begin{table}[!htp]
\centering
\caption{Results of the hyperparameter study for the FNO-based architecture. Each investigated architecture is characterized by a number of Fourier layers $n_\mathrm{f}$, a width of the latent representation $n_\mathrm{w}$ and a number of modes $n_\mathrm{m}$ for truncation of the layers' Fourier series. Our observations reveal that the performance on the validation set does not show relevant improvement for $n_\mathrm{w}>32$, and that $n_\mathrm{f}=3$ lead to slightly better performance than the larger number of layers. The performance, measured by the mean nL2 value on validation set, displays a little more distinct variation among architectures than the U-Net before. Nevertheless, neighbouring parameter combination of the FNO result in almost the same performance,  suggesting that the FNO, is also not overly sensitive to hyperparameter changes. We select the best model according to the smallest mean validation sets nL2 highlighted in blue as our $\mathcal{M}_{\mathrm{F},1}$.}
\begin{scriptsize}
\begin{tabular}{lccrrcccccc}
\toprule
\toprule
\multicolumn{3}{c}{FNO hyperparameters} & \multicolumn{2}{c}{} & \multicolumn{2}{c}{$\mathrm{nL2}$} & \multicolumn{2}{c}{$\frac{\mathrm{nL2}_\mathrm{shad}}{\mathrm{nL2}_\mathrm{vis}}$}& \multicolumn{2}{c}{SSP} \\
\cmidrule(rl){1-3} \cmidrule(rl){6-7} \cmidrule(rl){8-9} \cmidrule(rl){10-11}
layers $n_\mathrm{f}$ & modes $n_\mathrm{m}$ & width $n_\mathrm{w}$ & \#weights & epochs & train & val & train & val & train & val  \\ 
\midrule
\midrule
 
\multirow{4}{*}{3} & \multirow{4}{*}{32} 
      & 16 &   52,321 & 793 & 0.316 & 0.334 & 1.380 & 1.437 & 0.163 & 0.173 \\
    & & 32 &  204,225 & 790 & 0.262 & 0.313 & 1.338 & 1.520 & 0.134 & 0.160 \\
    & & 48 &  455,969 & 540 & 0.216 & 0.296 & 1.371 & 1.685 & 0.110 & 0.150 \\
    & & 64 &  807,553 & 500 & 0.214 & 0.296 & 1.362 & 1.681 & 0.109 & 0.151\vspace{0.12cm}\\

\multirow{4}{*}{3} & \multirow{4}{*}{40} 
      & 16 &   64,609 & 791 & 0.284 & 0.306 & 1.469 & 1.547 & 0.145 & 0.157 \\
    & & 32 &  253,377 & 703 & 0.229 & 0.290 & 1.402 & 1.654 & 0.116 & 0.148 \\
    & & 48 &  566,561 & 480 & 0.212 & 0.291 & 1.380 & 1.701 & 0.108 & 0.148 \\
    & & 64 & 1,004,161 & 395 & 0.207 & 0.288 & 1.390 & 1.719 & 0.105 & 0.146\vspace{0.12cm}\\

\multirow{4}{*}{3} & \multirow{4}{*}{48} 
      & 16 &   76,897 & 788 & 0.272 & 0.296 & 1.481 & 1.573 & 0.139 & 0.152 \\
    & & 32 &  302,529 & 605 & 0.219 & 0.279 & 1.419 & 1.675 & 0.111 & 0.142 \\
    & & 48 &  677,153 & 342 & 0.216 & 0.285 & 1.373 & 1.661 & 0.110 & 0.145 \\
    & & 64 & 1,200,769 & 258 & 0.214 & 0.286 & 1.360 & 1.642 & 0.109 & 0.146\vspace{0.12cm}\\

\multirow{4}{*}{3} & \multirow{4}{*}{56} 
      & 16 &   89,185 & 787 & 0.243 & 0.270 & 1.602 & 1.704 & 0.124 & 0.137 \\
    & & 32 &  351,681 & 714 & 0.202 & 0.265 & 1.499 & 1.817 & 0.102 & 0.135 \\
    & & 48 &  787,745 & 397 & 0.208 & 0.265 & 1.511 & 1.767 & 0.105 & 0.134 \\
    & & 64 & 1,397,377 & 307 & 0.204 & 0.267 & 1.475 & 1.753 & 0.104 & 0.136\vspace{0.12cm}\\

\multirow{4}{*}{3} & \multirow{4}{*}{64} 
      & 16 &  101,473 & 798 & 0.237 & 0.265 & 1.644 & 1.767 & 0.102 & 0.135 \\
    & & 32 & \cellcolor{DynCyan!40} 400,833 & \cellcolor{DynCyan!40} 534 & \cellcolor{DynCyan!40} 0.199 & \cellcolor{DynCyan!40} 0.256 & \cellcolor{DynCyan!40} 1.561 & \cellcolor{DynCyan!40} 1.837 & \cellcolor{DynCyan!40} 0.101 & \cellcolor{DynCyan!40} 0.130 \\
    & & 48 &  898,337 & 349 & 0.199 & 0.257 & 1.560 & 1.844 & 0.101 & 0.130 \\
    & & 64 & 1,593,985 & 276 & 0.190 & 0.261 & 1.501 & 1.822 & 0.096 & 0.133\vspace{0.12cm}\\

\multirow{4}{*}{3} & \multirow{4}{*}{72} 
      & 16 &  113,761 & 742 & 0.234 & 0.267 & 1.588 & 1.750 & 0.119 & 0.136 \\
    & & 32 &  449,985 & 600 & 0.197 & 0.257 & 1.575 & 1.887 & 0.100 & 0.131 \\
    & & 48 & 1,008,929 & 367 & 0.189 & 0.258 & 1.568 & 1.889 & 0.096 & 0.132 \\
    & & 64 & 1,790,593 & 244 & 0.187 & 0.258 & 1.523 & 1.851 & 0.095 & 0.131\vspace{0.05cm} \\
\midrule
   
\multirow{4}{*}{4} & \multirow{4}{*}{32} 
      & 16 &   68,977 & 789 & 0.281 & 0.316 & 1.386 & 1.520 & 0.144 & 0.162 \\
    & & 32 &  270,817 & 520 & 0.220 & 0.296 & 1.355 & 1.668 & 0.112 & 0.151 \\
    & & 48 &  605,777 & 394 & 0.190 & 0.289 & 1.358 & 1.760 & 0.097 & 0.147 \\
    & & 64 & 1,073,857 & 233 & 0.185 & 0.292 & 1.303 & 1.711 & 0.094 & 0.148\vspace{0.12cm}\\

\multirow{4}{*}{4} & \multirow{4}{*}{40} 
      & 16 &   85,361 & 787 & 0.250 & 0.291 & 1.469 & 1.658 & 0.128 & 0.148 \\
    & & 32 &  336,353 & 608 & 0.193 & 0.281 & 1.397 & 1.817 & 0.098 & 0.143 \\
    & & 48 &  753,233 & 331 & 0.188 & 0.283 & 1.360 & 1.771 & 0.095 & 0.144 \\
    & & 64 & 1,336,001 & 167 & 0.206 & 0.293 & 1.314 & 1.638 & 0.105 & 0.149\vspace{0.12cm}\\

\multirow{4}{*}{4} & \multirow{4}{*}{48} 
      & 16 &  101,745 & 720 & 0.245 & 0.289 & 1.408 & 1.612 & 0.125 & 0.148 \\
    & & 32 &  401,889 & 327 & 0.212 & 0.282 & 1.365 & 1.662 & 0.103 & 0.143 \\
    & & 48 &  900,689 & 185 & 0.205 & 0.282 & 1.355 & 1.676 & 0.104 & 0.144 \\
    & & 64 & 1,598,145 & 431 & 0.144 & 0.282 & 1.324 & 1.785 & 0.073 & 0.142\vspace{0.12cm}\\

\multirow{4}{*}{4} & \multirow{4}{*}{56} 
      & 16 &  118,129 & 703 & 0.216 & 0.265 & 1.533 & 1.767 & 0.110 & 0.135 \\
    & & 32 &  467,425 & 518 & 0.172 & 0.269 & 1.420 & 1.873 & 0.087 & 0.136\\
    & & 48 & 1,048,145 & 285 & 0.171 & 0.269 & 1.406 & 1.834 & 0.087 & 0.137 \\
    & & 64 & 1,860,289 & 156 & 0.177 & 0.269 & 1.385 & 1.778 & 0.089 & 0.136\vspace{0.12cm}\\

\multirow{4}{*}{4} & \multirow{4}{*}{64} 
      & 16 &  134,513 & 784 & 0.204 & 0.258 & 1.555 & 1.843 & 0.103 & 0.131 \\
    & & 32 &  532,961 & 270 & 0.186 & 0.258 & 1.497 & 1.846 & 0.095 & 0.131 \\
    & & 48 & 1,195,601 & 149 & 0.187 & 0.260 & 1.476 & 1.831 & 0.094 & 0.132 \\
    & & 64 & 2,122,433 & 114 & 0.176 & 0.259 & 1.450 & 1.826 & 0.089 & 0.131\vspace{0.12cm}\\

\multirow{4}{*}{4} & \multirow{4}{*}{72} 
      & 16 &  150,897 & 740 & 0.189 & 0.256 & 1.528 & 1.903 & 0.095 & 0.130 \\
    & & 32 &  598,497 & 262 & 0.180 & 0.257 & 1.496 & 1.903 & 0.091 & 0.130 \\
    & & 48 & 1,343,057 & 181 & 0.168 & 0.257 & 1.454 & 1.879 & 0.085 & 0.130 \\
    & & 64 & 2,384,577 & 121 & 0.169 & 0.259 & 1.406 & 1.817 & 0.086 & 0.132\vspace{0.05cm} \\
\bottomrule
\bottomrule
\end{tabular}
\end{scriptsize}
\label{tab:hyperparametersFNO}
\vspace{1.2cm}
\end{table}

\section{Loss curves}
\label{sec:Appendix Loss Curves}

\begin{figure}[!htp]
     \centering
     \begin{subfigure}[b]{0.495\textwidth}
         \centering
         \includegraphics[width=\textwidth]
         {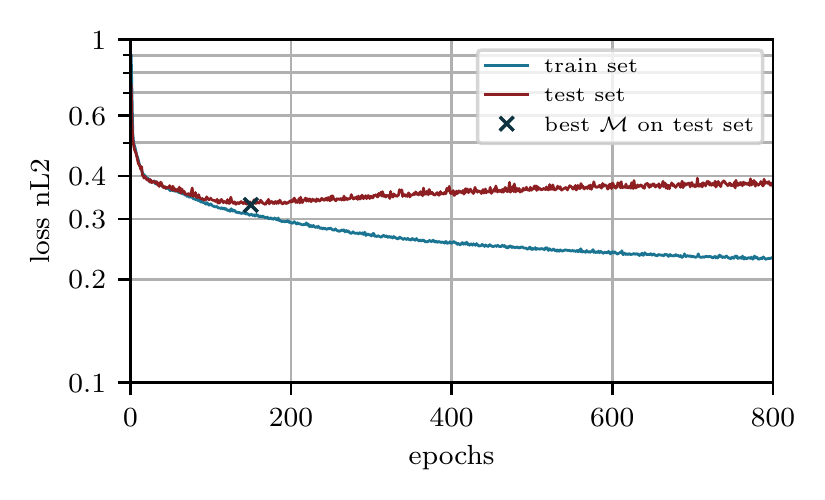}
         \caption{}
         \label{fig:loss_UNet_1_snap}
     \end{subfigure}
     \hfill
     \begin{subfigure}[b]{0.495\textwidth}
         \centering
         \includegraphics[width=\textwidth]{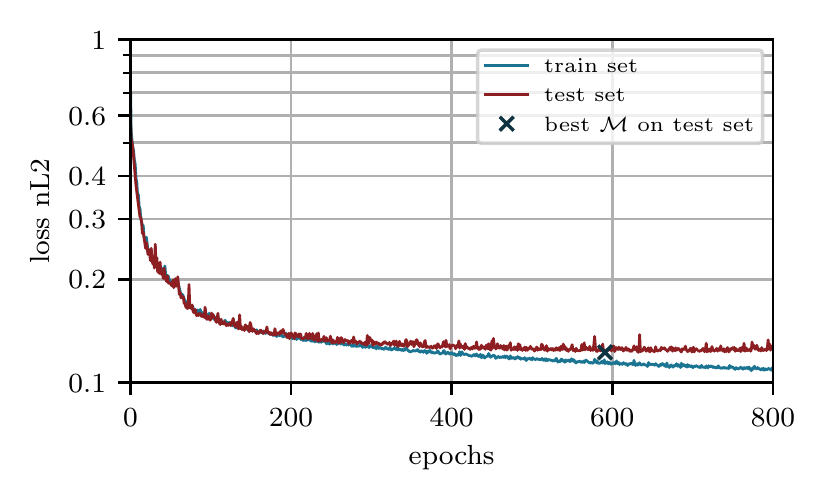}
         \caption{}
         \label{fig:loss_UNet_10_snap}
     \end{subfigure}
     \caption{Loss curves for training of the U-Net-based model. Subfigure (a) depicts the loss of model $\mathcal{M}_{\mathrm{U},1}$ trained with one one snapshot $n_\mathrm{s}=1$ in the radar input, where the best performance $\mathrm{nL2}=0.329$ on test set for model evaluation is reached after 150 epochs. Afterwards the model would tend to overfit the training data. Subfigure (b) depicts model $\mathcal{M}_{\mathrm{U},10}$ trained with $n_\mathrm{s}=10$ instead, which strongly increases performance, resulting in $\mathrm{nL2}=0.123$ after 592 epochs of training. }
\end{figure}

\begin{figure}[ht]
     \centering
     \begin{subfigure}[b]{0.495\textwidth}
         \centering
         \includegraphics[width=\textwidth]
         {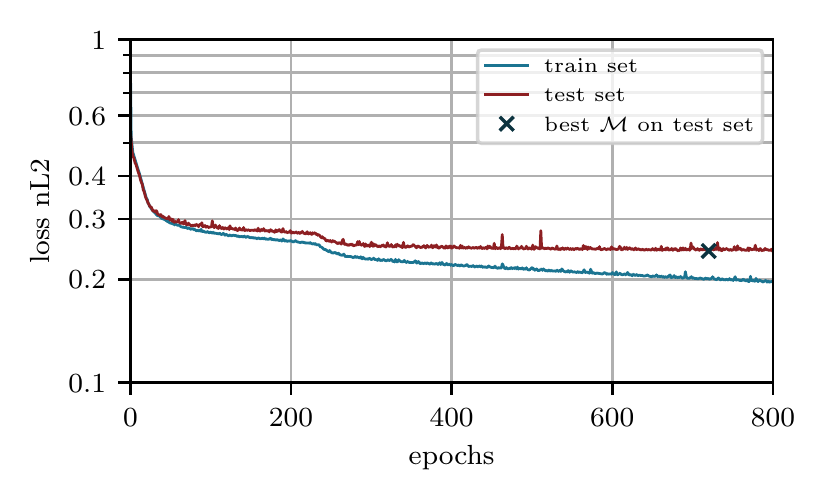}
         \caption{}
         \label{fig:loss_FNO_1_snap}
     \end{subfigure}
     \hfill
     \begin{subfigure}[b]{0.495\textwidth}
         \centering
         \includegraphics[width=\textwidth]{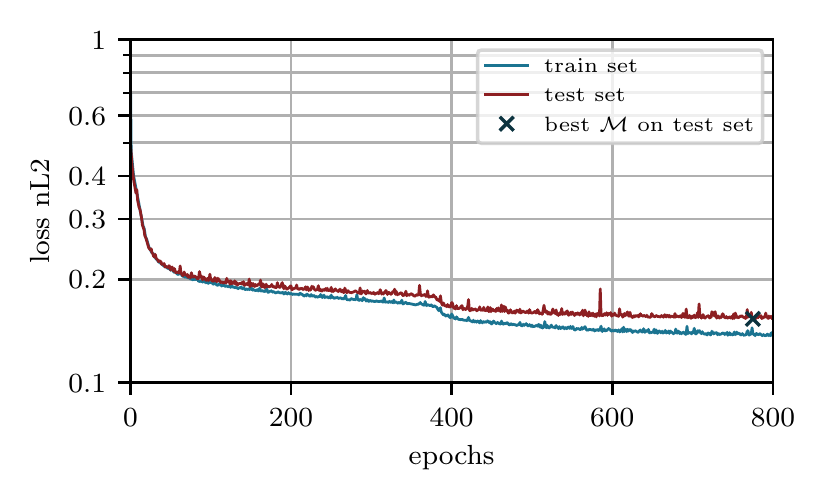}
         \caption{}
         \label{fig:loss_FNO_9_snap}
     \end{subfigure}
     \caption{Loss curves for training of the FNO-based model. Subfigure (a) depicts the loss of model $\mathcal{M}_{\mathrm{F},1}$ trained with one one snapshot $n_\mathrm{s}=1$ in the radar input, where the best performance $\mathrm{nL2}=0.242$ on test set for model evaluation is reached after 721 epochs. Compared to the U-Net based model $\mathcal{M}_{\mathrm{U},1}$, $\mathcal{M}_{\mathrm{F},1}$ does not seem to be susceptible to overfitting . Subfigure (b) depicts model $\mathcal{M}_{\mathrm{F},9}$ trained with $n_\mathrm{s}=9$ instead, which increases performance, resulting in $\mathrm{nL2}=0.153$ after 776 epochs of training. It can be expected that training beyond 800 epochs would further slightly increase the best performance on test set. }
\end{figure}

After determining appropriate hyperparameters for the U-Net-based and FNO-based models in \ref{sec:Appendix-Hyperparameters}, the train and validation data from the four-fold cross-validation were merged. This combined data set was then used to train the models $\mathcal{M}_{\mathrm{U}, n_\mathrm{s}}$ and $\mathcal{M}_{\mathrm{F}, n_\mathrm{s}}$, with one radar snapshot in each samples input ($n_\mathrm{s}=1$) or either $n_\mathrm{s}=9$ or $n_\mathrm{s}=10$ historical radar snapshots in each input. The performance evaluation of these models was conducted on the previously excluded test set of $N_\mathrm{test}=624$ samples. The loss curves depicted in Figures \ref{fig:loss_UNet_1_snap}-\ref{fig:loss_FNO_9_snap} illustrate the model performance and the impact of different values for $n_\mathrm{s}$ throughout the training epochs. Deviation between the train and test loss curves indicates overfitting, characterized by excessive adaptation to the training data, resulting in poor generalization to new samples. Consequently, the best models $\mathcal{M}$ were selected based on the lowest test loss within the 800 training epochs.


\clearpage

\section{Visualization of spectral representation}

\begin{figure}[!htp]
\centering
\includegraphics{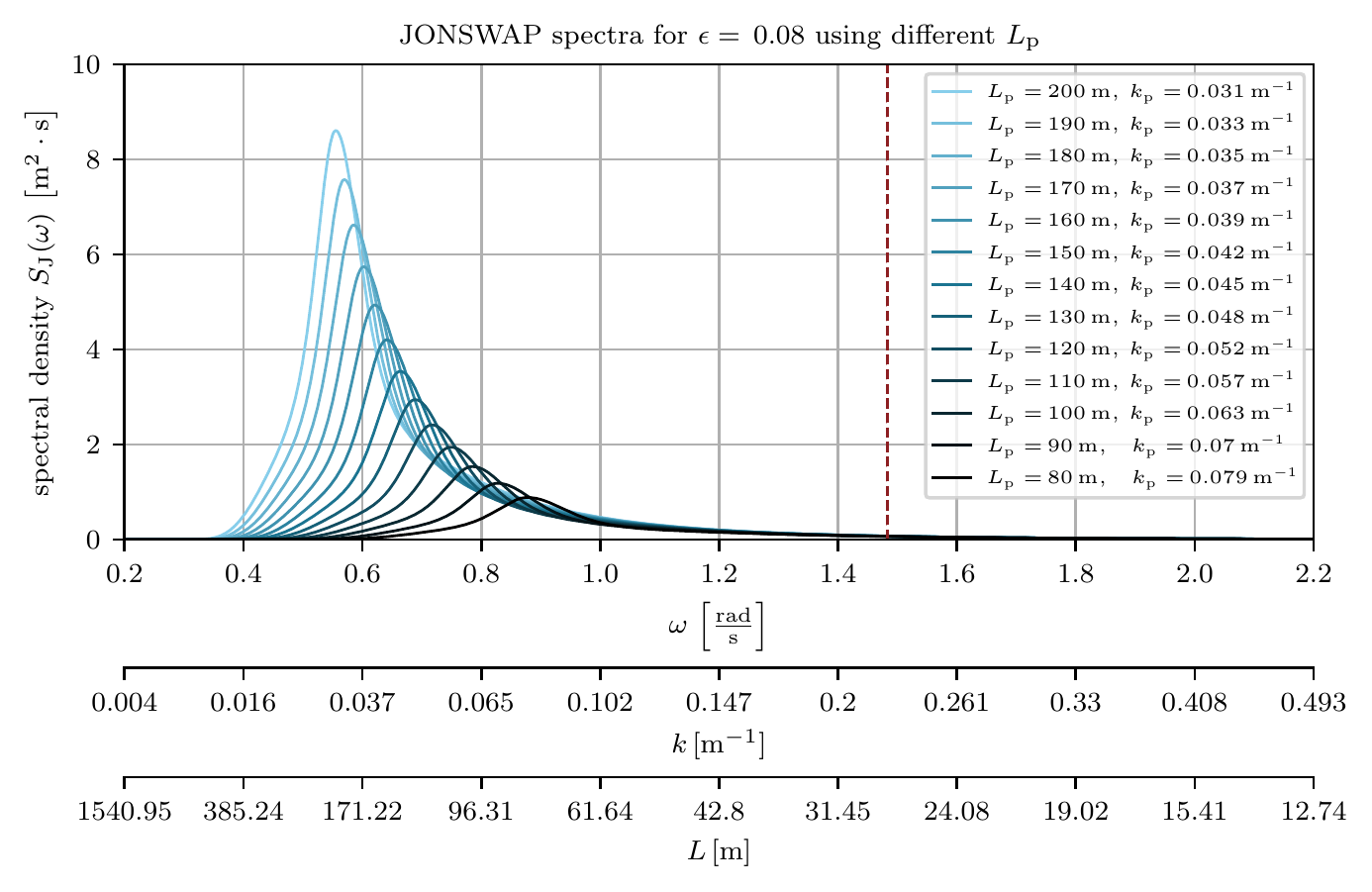}
\caption{JONSWAP spectra used in the data generation for one exemplary steepness value $\epsilon$, but varying peak wavelengths $L_\mathrm{p}=80-200 \, \mathrm{m}$. The shortest peak wavelength of $L_\mathrm{p}=80 \, \mathrm{m}$ corresponds to the highest peak wavenumber  of $k_\mathrm{p} = 0.079 \, \mathrm{m}^{-1}$. The filtering wavenumber of $k_\mathrm{filt}= n_\mathrm{m} \cdot \Delta k = 64 \cdot 0.00351 \, \mathrm{m}^{-1} = 0.2246 \, \mathrm{m}^{-1} $, which is indicated by the dotted red line and defined by the Fourier layers in this work, consequently does not truncate important wave components in our data set-up.}
\label{fig:JONSWAP_Lps}
\end{figure}

During the investigations on the FNO models (see Figure \ref{fig:FNO}), a concern arose regarding the chosen number of Fourier series modes $n_\mathrm{m}=64$ in the $R_i$-matrices, which might lead to the omission of significant frequency components in the wave data. To address this concern, we visualized the JONSWAP spectra employed to initialize the HOS wave simulation for a specific steepness value $\epsilon$ (since different $\epsilon=0.08$ just scale the amplitude of spectral density) and all peak wavelengths $L_\mathrm{p} \in \{80, 90, \hdots, 190, 200 \} \, \mathrm{m}$, each corresponding to a specific $\omega_\mathrm{p}$ and $k_\mathrm{p}$. Based on the findings depicted and explained in Figure \ref{fig:JONSWAP_Lps}, we conclude that this assumption is invalid.

\section*{Acknowledgements}
This work was supported by the Deutsche Forschungsgesellschaft (DFG - German Research Foundation) [project number 277972093: Excitability of Ocean Rogue Waves]

\section*{Declaration of interests}

The authors declare that they have no known competing financial interests or personal relationships that could have appeared to influence the work reported in this paper.

\section*{Declaration of Generative AI and AI-
assisted technologies in the writing process}

The manuscript was completely written by the authors. Once the manuscript was completed, the authors used ChatGPT in order to improve its grammar and readability. After using this tool, the authors reviewed and edited the content as needed and take full responsibility for the content of the publication.

\bibliography{bibliography}

\end{document}